\documentclass[aps,prl,twocolumn,reprint,superscriptaddress]{revtex4-2}
\usepackage[colorlinks=true, letterpaper=true, pdfstartview=FitV, linkcolor=blue, citecolor=blue, urlcolor=blue]{hyperref}
\usepackage{amsfonts}
\usepackage{amssymb}
\usepackage{amsmath}
\usepackage{graphicx}
\usepackage{dcolumn}
\usepackage{bm}
\usepackage{float}
\usepackage{setspace}
\usepackage{epsfig}
\usepackage{array}
\usepackage{braket}
\usepackage{multirow}
\usepackage[table,xcdraw]{xcolor}
\usepackage{utfsym}
\usepackage{booktabs} 
\usepackage{makecell}   
\usepackage{appendix}

\begin{document}

\title{Sliding ferroelectric control of unconventional magnetism in stacked bilayers}

\author{Yongqian Zhu}
\affiliation{Beijing National Laboratory for Condensed Matter Physics, Institute of Physics, Chinese Academy of Sciences, Beijing 100190, China}
\affiliation{University of Chinese Academy of Sciences, Chinese Academy of Sciences, Beijing 100190, China}

\author{Mingqiang Gu}%
\affiliation{State key laboratory of quantum functional materials, Department of Physics, and Guangdong Basic Research Center of Excellence for Quantum Science, Southern University of Science and Technology (SUSTech), Shenzhen 518055, China}

\author{Yuntian Liu}%
\affiliation{State key laboratory of quantum functional materials, Department of Physics, and Guangdong Basic Research Center of Excellence for Quantum Science, Southern University of Science and Technology (SUSTech), Shenzhen 518055, China}

\author{Xiaobing Chen}%
\affiliation{State key laboratory of quantum functional materials, Department of Physics, and Guangdong Basic Research Center of Excellence for Quantum Science, Southern University of Science and Technology (SUSTech), Shenzhen 518055, China}
\affiliation{China Quantum Science Center of Guangdong-Hong Kong-Macao Greater Bay Area (Guangdong), Shenzhen 518045, China}

\author{Yuhui Li}%
\affiliation{State key laboratory of quantum functional materials, Department of Physics, and Guangdong Basic Research Center of Excellence for Quantum Science, Southern University of Science and Technology (SUSTech), Shenzhen 518055, China}

\author{Shixuan Du}
\email{sxdu@iphy.ac.cn}
\affiliation{Beijing National Laboratory for Condensed Matter Physics, Institute of Physics, Chinese Academy of Sciences, Beijing 100190, China}
\affiliation{University of Chinese Academy of Sciences, Chinese Academy of Sciences, Beijing 100190, China}
\affiliation{Songshan Lake Materials Laboratory, Dongguan 523808, China}

\author{Qihang Liu}%
\email{liuqh@sustech.edu.cn}
\affiliation{State key laboratory of quantum functional materials, Department of Physics, and Guangdong Basic Research Center of Excellence for Quantum Science, Southern University of Science and Technology (SUSTech), Shenzhen 518055, China}
\affiliation{China Quantum Science Center of Guangdong-Hong Kong-Macao Greater Bay Area (Guangdong), Shenzhen 518045, China}

\begin{abstract}
The control of unconventional magnetism, which displays
ferromagnetism-like properties with compensated magnetization, has drawn
intense attention for advancing antiferromagnetic spintronics. Here,
through symmetry analysis, we propose a general stacking rule,
characterized by a connection operator linking two stacked bilayers, for
controlling unconventional magnetism via sliding ferroelectricity. Such
rule enables the simultaneous switching of both electric polarization
and nonrelativistic spin splitting or anomalous Hall effect in
altermagnets, a class of collinear unconventional magnets. By
comprehensively surveying the 80 layer groups, we identify all the
stacking orders that allow for such two types of simultaneous switching.
Furthermore, we extend the stacking rule to collinear compensated
ferrimagnets, where the opposite-spin sublattices are not connected by
any symmetry operator, yet the net magnetization remains zero. Combined
with first-principles calculations, we demonstrate the sliding
ferroelectric control of spin polarization and anomalous Hall effect in
the altermagnetic AgF\textsubscript{2} and
Fe\textsubscript{2}MoSe\textsubscript{4} bilayers. Our work provides a
symmetry strategy for achieving ferroelectric control of unconventional
magnetism in bilayer systems and opens avenues for exploring new types
of magnetoelectric coupling.
\end{abstract}
\maketitle

\emph{Introduction}---Unconventional magnetism is broadly characterized by compensated
magnetization yet exhibiting ferromagnetism-like properties, including
spin splitting, anomalous Hall effect (AHE), quantum geometry,
topological magnons, etc \cite{1000}. Recent advances
in classifying unconventional magnets according to spin-group symmetry
have expanded the current understanding of magnetism
 \cite{1042, 743, 773, 669, 147, 580, 255, 898, 671}. A
prominent example is altermagnetism, a type of collinear antiferromagnetism
(AFM) that displays nonrelativistic spin splitting in momentum space
 \cite{897, 658, 894, 848, 580, 255, 693}. The spin
splitting originates from the collinear magnetic order rather than from
relativistic spin-orbit coupling (SOC). Such magnetic order-induced spin
polarization enables various spintronic applications, such as
spin-polarized currents, spin-to-charge conversion, spin torques, and
magnetoresistance \cite{983, 793, 986, 658, 870, 987,
957, 817, 255}. Another representative category of unconventional
magnetism is the AFM exhibiting the anomalous Hall effect (AHE)
 \cite{368, 926, 212, 599, 927, 480},  which enables the
electrical read-out of the magnetic state. In collinear and coplanar
AFMs, the occurrence of the AHE necessarily requires SOC, while for
noncoplanar AFM, it can originate solely from the magnetic order
 \cite{679}.

In AFM memories, controlling the two key properties of unconventional
magnetism (i.e., spin splitting and AHE), typically by means of
manipulating AFM moments by a spin torque, opens new possibilities for
information writing \cite{370, 483}. While
dissipationless spin-orbit torque can be driven by an in-plane electric
field \cite{1059, 208, 1053}, manipulating AFM moments
usually suffers from large energy dissipation as spin torque requires
electric current in most cases \cite{845, 483, 1006,
891, 893, 1005}. Alternatively, sliding ferroelectricity provides an
energy-efficient way \cite{1058} for controlling
unconventional magnetism through a gate voltage, e.g., switching the
spin polarization in altermagnets \cite{730}. Stacking
has been previously demonstrated as an effective approach for inducing
either sliding ferroelectricity or altermagnetism in a wide range of
two-dimensional (2D) materials \cite{866, 128, 40, 864,
882, 861,1088}. However, a universal and efficient strategy for designing
materials with coupled sliding ferroelectricity and multiple facets of
unconventional magnetism is still lacking.

In this Letter, we propose a general symmetry rule for controlling
unconventional magnetism via sliding ferroelectricity in bilayer
systems. Such a symmetry rule can easily predict whether electric
polarization and unconventional magnetism are coupled in stacked
bilayers, based solely on the crystallographic layer group of their
constituent monolayers, the stacking operations and magnetic
configurations of the bilayers. Combined with first-principles
calculations, we demonstrate that in altermagnetic bilayer
AgF\textsubscript{2}, spin polarization and AHE can be independently
controlled via sliding ferroelectricity. In contrast, simultaneous
control of both spin polarization and AHE is achieved in altermagnetic
bilayer Fe\textsubscript{2}MoSe\textsubscript{4} and compensated
ferrimagnetic bilayer CrI\textsubscript{3}. Such simultaneous control
provides an effective route to detect spin polarization switching
through AHE measurements.

\emph{Symmetry rules for sliding ferroelectricity}---For stacked bilayers, the existence of spontaneous electric polarization
or unconventional magnetism is theoretically governed by their
symmetries established on different group frameworks
 \cite{128, 882, 304, 1009, 1007, 1011}. The switching
of these properties between two bilayer configurations is determined by
an operator that connects them \cite{995, 848, 849, 343}, referred to as the connection operator. Obviously, the connection
operator determines the coupling of ferroelectricity and unconventional
magnetism. Based on these symmetry rules, we identify ferroelectric
bilayers that can control unconventional magnetism as follows.

\begin{figure}
    \includegraphics[width=8.6cm]{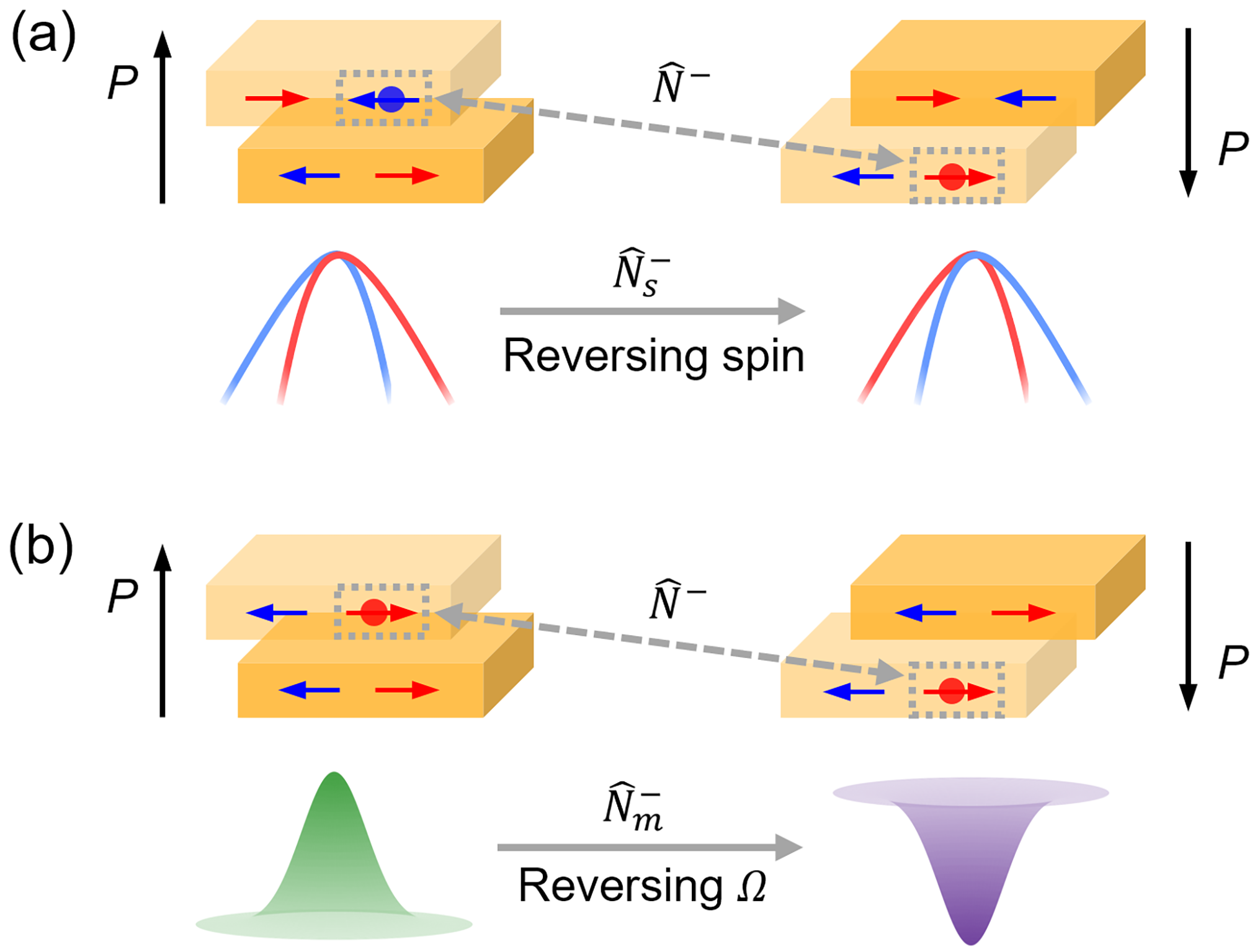}
    \caption{Schematic diagrams of (a) configurations (upper) and bands
    (lower), (b) configurations (upper) and Berry curvatures (lower) of
    ferroelectric altermagnetic bilayers. The red and blue colors of arrows
    (lines) denote the opposite magnetic moments (spins). The black arrow
    denotes \(P_{\ }\). For the two stacked bilayers in (a) or (b), one can
    be transformed into the other through interlayer sliding. The sliding is
    equivalent to applying a \({\widehat{N}}_{\ }^{-}\) transformation,
    which maps the top (bottom) layer of one bilayer to the bottom (top)
    layer of the other. The solid circles in grey dotted line boxes denote
    atoms connected by \({\widehat{N}}_{\ }^{-}\), which have opposite spins
    in (a) while the same spin in (b).}
    \label{Fig.1}
\end{figure}

We first screen out all the stacked bilayers with sliding
ferroelectricity across all 80 layer groups. A bilayer system (\(B\))
consisting of a bottom layer (\(S\)) and a top layer (\(S'\)) can be
expressed as \(B = S + S' = S + \widehat{O}S\), where
\(\widehat{O} = \left\{ O \middle| \boldsymbol{t}_{o} \right\}\) is a
stacking operator transforming \(S\) into \(S'\)
 \cite{128}. Here, \(O\) is the rotational part, and
\(\boldsymbol{t}_{o}\) is the translational part. A stacking operator
\(\widehat{O}\) specifies a stacking order. We then consider a
ferroelectric bilayer with the same lateral unit cell size as its
constituent monolayers, possessing an out-of-plane electric polarization
component (\(P_{\ }\)). The symmetry operator (\(\widehat{R}\)) of a
bilayer system and the connection operator (\(\widehat{N}\)) between two
bilayer systems {[}see Fig. \ref{Fig.1}(a){]} can be divided into two classes,
\({\widehat{Q}}_{\ }^{-}\) and \({\widehat{Q}}_{\ }^{+}\)
(\({\widehat{Q}}_{\ } = \widehat{R},\ {\widehat{N}}_{\ }\)), which can
and cannot reverse \(P_{\ }\), respectively. For each layer group of the
constituent monolayer, we identify all the stacking orders with only
\({\widehat{R}}_{\ }^{+}\) symmetries that permit \(P_{\ }\)
 \cite{128}. Then, we further screen out those with
switchable \(P_{\ }\) through interlayer sliding (i.e., sliding
ferroelectricity). For two bilayer configurations, opposite \(P_{\ }\)
requires a connection operator \(\widehat{N} = {\widehat{N}}_{\ }^{-}\),
as shown in Fig. \ref{Fig.1}. By solving the equation set for
\({\widehat{N}}_{\ }^{-}\) across 80 layer groups, we identify all the
stacking orders with switchable \(P_{\ }\), as detailed in Sec. 1 \cite{1013}. Table \ref{table1} presents an
example for layer groups No. 14-18 of the monolayer. A bilayer permits
\(P_{\ }\) when stacked with \(O = m_{001}\) and \(\boldsymbol{t}_{o}\)
along the high symmetry line GB or CA. \(P\) can be switched when
changing the stacking order from \(\boldsymbol{t}_{o}\) to
\(- N^{-}\boldsymbol{t}_{o} + \boldsymbol{t}_{0}\) through interlayer sliding,
where \(\boldsymbol{t}_{0}\) represents the pure translational symmetry of
the constituent monolayer.

\renewcommand{\arraystretch}{1.5} \begin{table}[b]
    \fontsize{6}{7.2}\selectfont
        \caption{Stacking configuration of layer groups No. 14-18 for the
        sliding ferroelectric control of spin polarization and AHE. \(G_{S}\)
        (\(G_{B}\)) denotes the layer group of the monolayer (bilayer);
        \(p2/m11\) and \(p2_{1}/m11\) in \(G_{S}\) correspond to \(pm11\) in
        \(G_{B}\), \(p2/b11\) and \(p2_{1}/b11\) in \(G_{S}\) correspond to
        \(pb11\) in \(G_{B}\), and \(c2/m11\) in \(G_{S}\) corresponds to
        \(cm11\) in \(G_{B}\). \(\boldsymbol{t}_{o}\) is represented by high
        symmetry lines \(GB\) and \(CA\) in the rectangular lattice, with
        \(G = (0,\ 0)\), \(A = (\frac{1}{2},\ 0)\),
        \(B = (0,\ \ \frac{1}{2}\ )\) and \(C = (\frac{1}{2},\ \ \frac{1}{2})\).
        For \(N_{s}^{-}\) (\(N_{m}^{-}\)), only the operators capable of
        switching the spin polarization (AHE) are listed. The candidate
        monolayers are screened from Ref. \cite{751}.}
    
        \begin{ruledtabular} %
            
               \begin{tabular}{ccccccc}
                $G_S$ (No.) & $\{O|\boldsymbol{t}_o\}$ & $G_B$ & $N^-$ & $N_s^-$ & $N_m^-$ & Candidates \\
                \hline
                \makecell{ $p2/m11(14)$ \\ $p2_1/m11(15)$ \\ $p2/b11(16)$ \\ $p2_1/b11(17)$ \\ $c2/m11(18)$ } &
                $\{m_{001}|$GB,CA\} &
                \makecell{ $pm11$ \\ $pb11$ \\ $cm11$ } &
                \makecell{ $m_{001}$ \\ 2$_{010}$ } &
                \makecell{ $\{2_\perp||m_{001}\}$ \\ $\{1||2_{010}\}$ } &
                \makecell{ $m'_{001}$ \\ 2$_{010}$ } &
                \makecell{ AgF$_2$, RuF$_4$,\\ VF$_4$, OsF$_4$ } \\
    
                \end{tabular} 
        \end{ruledtabular}
        \label{table1}
    \end{table}

\emph{Sliding ferroelectric control of spin polarization}---Having obtained all stacked bilayers with switchable \(P_{\ }\), we turn
to identify those with coupled ferroelectricity and unconventional
magnetism. We first consider ferroelectric control of spin polarization
in altermagnets. For 2D altermagnets, the symmetry \(\widehat{R}\), that
connects two sublattices with opposite spins, cannot be a translation
\(\boldsymbol{t}\), inversion \(\overline{1}\), rotation \(2_{001}\), or any
of their combinations \cite{866, 751, 756}. Because
the considered ferroelectric bilayer contains only \({\widehat{R}}^{+}\)
symmetry, the requirement is reduced to \(\widehat{R}\) not being
\(\boldsymbol{t}\), \(2_{001}\) or their combinations. For example,
altermagnetism is allowed in the ferroelectric bilayer with layer group
\(pb11\) (see Table \ref{table1}). Thus, we identify all the stacking orders
allowing for altermagnetism from screened ferroelectric bilayers, as
highlighted in Tables S1-S3
 \cite{1013}. It is worth noting that the ferroelectric
altermagnetic bilayers discussed here are fundamentally different from
previously reported altermagnetic bilayers with
\(\widehat{R} = {\widehat{R}}^{-}\) (e.g., A-type AFM)
 \cite{882, 864, 861}, where \({\widehat{R}}^{-}\)
enforces a vanishing \(P\). In contrast, the \({\widehat{R}}^{+}\)
symmetry in a bilayer system considered here originates from the
constituent monolayer, a collinear antiferromagnet with different
sublattices connected by \({\widehat{R}}^{+}\).

Then, we further identify the ferroelectric altermagnetic bilayers with
switchable spin polarization. We introduce the connection operator
\({\widehat{N}}_{s}\) {[}Fig. \ref{Fig.1}(a){]}, which is an operator of the spin
space group \cite{147, 580, 743, 773, 898}. The
interlayer magnetic coupling includes two types: the Néel vectors of the
two constituent monolayers are aligned in the opposite or the same
direction {[}Figs. \ref{Fig.1}(a) and \ref{Fig.1}(b){]}. We assume that the collinear
magnetic moment of each ion remains unchanged under interlayer sliding.
Consequently, for the two atoms with opposite spins connected by
\({\widehat{N}}_{\ }^{–}\) {[}Fig. \ref{Fig.1}(a){]}, considering spin rotation,
they are connected by \({\widehat{N}}_{s}^{-}\), where the rotational
part is given by \(N_{s}^{-} = \{ 2_{\bot}||N^{-}\}\). Here the
\(2_{\bot}\) indicates a twofold spin rotation along the axis
perpendicular to the collinear magnetic moments. Under
\(N_{s}^{-} = \{ 2_{\bot}||N^{-}\}\) operation, the spin polarization
distribution with respect to the momentum \(s^\text{I}(\boldsymbol{k})\)
transforms into
\(s^{\text{\uppercase\expandafter{\romannumeral2}}}\left( \boldsymbol{k} \right) = N_{s}^{-}s^\text{I}\left( \boldsymbol{k} \right) = - s^\text{I}(\left( N^{-} \right)^{- 1}\boldsymbol{k})\),
where the superscripts I and \text{\uppercase\expandafter{\romannumeral2}} denote the two
ferroelectric states connected by \({\widehat{N}}_{s}^{-}\). Taking
\(N^{-} = m_{001}\) as an example, since
\(\left( m_{001} \right)^{- 1}\boldsymbol{\ k = k}\) for any \(\boldsymbol{k}\)
point, the spin polarization across the entire Brillouin zone can be
reversed under ferroelectric switching, i.e.,
\(s^{\text{\uppercase\expandafter{\romannumeral2}}}\left( \boldsymbol{k} \right) = - s^\text{I}(\boldsymbol{k})\), as
shown in Fig. \ref{Fig.1}(a) and Table \ref{table1}. Consequently, the connection operator
\(N_{s}^{-}\), which depends on \(N^{-}\) and the type of interlayer
magnetic coupling, determines the spin polarization reversal. We find
that for all the ferroelectric altermagnetic bilayers connected by
\(N^{-}\) screened above, at least one type of interlayer magnetic
coupling enables spin polarization reversal under ferroelectric
switching.

\emph{Sliding ferroelectric control of anomalous Hall effect}---Now we discuss the ferroelectric control of another facet of
unconventional magnetism, i.e., AHE in AFM. Distinct from spin
splitting, the symmetry requirement for AHE in 2D systems is that the
sign of Berry curvature (\(\Omega\)) remains unchanged under any
symmetry operation. We consider AHE in 2D altermagnets, indicating the
necessity of SOC and the framework of magnetic group, a specific
subgroup of spin group. To achieve AHE in antiferromagnets with
symmetry-enforced zero magnetization, the magnetic configuration must
not align along the out-of-plane direction, otherwise the symmetry
operations connecting the opposite-spin sublattices will reverse the
sign of Berry curvature and thus enforce a zero anomalous Hall
conductivity \cite{465}. We consider the in-plane
magnetic configuration. Because \(2_{001}\) symmetry of a 2D altermagnet
only connects the same-spin sublattice, it will result in a
\(2_{001}'\) symmetry in the magnetic point group for the in-plane
configuration, leading to a vanishing AHE. Hence, any altermagnetic
bilayers with \(2_{001}\) point-group symmetry should be ruled out. We
screen out all the stacking orders permitting AHE from the ferroelectric
altermagnetic bilayers, as highlighted in Tables S1-S3 \cite{1013}.

We further identify ferroelectric altermagnetic bilayers with
controllable AHE, which requires a sign reversal of Berry curvature
under ferroelectric switching {[}see Fig. \ref{Fig.1}(b){]}. When considering SOC,
the connection operator belongs to magnetic space groups and is denoted
as \({\widehat{N}}_{m}^{-}\). The collinear in-plane magnetic
configuration imposes a constraint that
\(N_{m}^{-} \in \{ m_{001},m_{001}',2_{\alpha},2_{\alpha}'\}\),
indicating that \(N_{\ }^{-} \in \{ m_{001},2_{\alpha}\}\), where
\(2_{\alpha}\) denotes a twofold rotation with the rotational axis along
in-plane \(\alpha\) direction. The sign reversal of Berry curvature
requires \(N_{m}^{-} \in \{ m_{001}',2_{\alpha}\}\) (see Table \ref{table1}).
This condition can be satisfied for all the above-screened ferroelectric
altermagnetic bilayers with AHE under an appropriate magnetic
configuration. Specifically, \(N_{m}^{-} = m_{001}'\) requires that
the two atoms connected by \(N^{-} = m_{001}\) have the same spin {[}see
Fig. \ref{Fig.1}(b){]}. On the other hand, \(N_{m}^{-} = 2_{\alpha}\) requires the
two atoms connected by \(N^{-} = 2_{\alpha}\) have the same (opposite)
spins when the spin is parallel (perpendicular) to the rotational axis
of \(2_{\alpha}\).

It is worth noting that AHE switching is not necessarily accompanied by
a spin polarization switching in momentum space. According to our
symmetry rules, for the stacking orders connected only by
\(N^{-} = 2_{\alpha}\), AHE and spin polarization can be switched
simultaneously under ferroelectric switching. We mark all the stacking
orders that allow for a simultaneous ferroelectric switching of both AHE
and spin polarization, as highlighted in Tables S1-S3 \cite{1013}.

\begin{figure}
	\includegraphics[width=8.6cm]{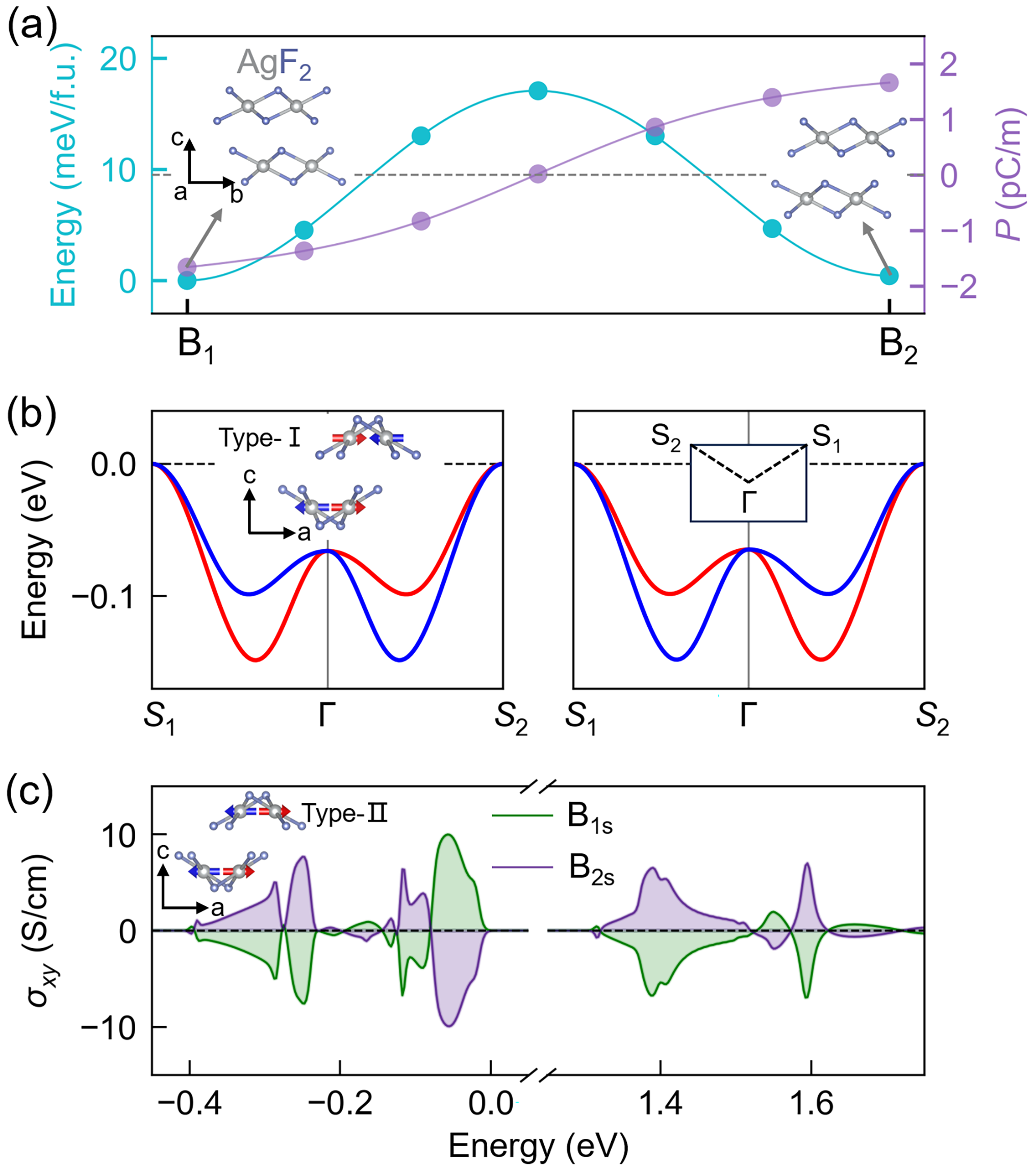}
	\caption{(a) The energy barrier and the evolution of out-of-plane
polarization along the sliding path between AgF\textsubscript{2}
bilayers B\textsubscript{1} and B\textsubscript{2}. The inset shows
configurations of B\textsubscript{1} and B\textsubscript{2}. (b) The two
highest occupied bands of B\textsubscript{1} (left) and
B\textsubscript{2} (right), calculated without considering SOC. The red
and blue colors denote opposite spins. The Fermi level is set to zero.
The left inset shows the type-\(\text{I}\) magnetic configuration, where
red and blue arrows denote opposite magnetic moments. The right inset
displays the Brillouin zone, where the high symmetry points
\(S_{1} = (\frac{1}{2},\frac{1}{2})\) and
\(S_{2} = ( - \frac{1}{2},\frac{1}{2})\). (c) The anomalous Hall
conductivities of B\textsubscript{1s} and B\textsubscript{2s},
calculated with considering SOC. The valence band maximum is set to
zero. The inset shows the type-\(\text{\uppercase\expandafter{\romannumeral2}}\) magnetic configuration,
where red and blue arrows denote the magnetic moment along
\(\lbrack 100\rbrack\) and \(\lbrack\overline{1}00\rbrack\) directions,
respectively.}
	\label{Fig.2}
\end{figure}

\emph{Sliding ferroelectric control of
unconventional magnetism in altermagnetic bilayer
AgF\textsubscript{2} and Fe\textsubscript{2}MoSe\textsubscript{4}}---We now apply the above symmetry analysis to realistic materials.
Monolayer candidate materials that enable ferroelectric control of spin
polarization and AHE via bilayer stacking, including
AgF\textsubscript{2} \cite{751},  MF\textsubscript{4}
(M =V, Ru, Os) \cite{751}, 
V\textsubscript{2}ClBrI\textsubscript{2}O\textsubscript{2}
\cite{751},  Fe\textsubscript{2}MX\textsubscript{4} (M
= Mo, W; X = S, Se, Te) \cite{1060}, 
MnPSe\textsubscript{3} \cite{205} and
FeBr\textsubscript{3} \cite{751},  are listed in Table
\ref{table1} and Tables S1-S3 \cite{1013}. Among these, stacked
bilayer AgF\textsubscript{2}, MF\textsubscript{4} (M =V, Ru, Os), and
V\textsubscript{2}ClBrI\textsubscript{2}O\textsubscript{2} allow for the
ferroelectric control of spin polarization or AHE independently, whereas
the others enable the ferroelectric simultaneous control of both spin
polarization and AHE. We take two representative monoalyers,
AgF\textsubscript{2} and Fe\textsubscript{2}MoSe\textsubscript{4}, as
illustrative examples.

The bulk phase of AgF\textsubscript{2} has been synthesized
experimentally \cite{901, 781}. Monolayer
AgF\textsubscript{2} is a \emph{d}-wave altermagnetic candidate with the
crystallographic layer group \(p2_{1}/b11\) (No. 17)
\cite{793, 751}. The sublattices with opposite spins
are connected by symmetries \(\{ m_{100}|\frac{1}{2},\frac{1}{2}\}\) and
\(\{ 2_{100}|\frac{1}{2},\frac{1}{2}\}\). Although ferroelectricity is
prohibited in monolayer AgF\textsubscript{2} due to the inversion
symmetry, according to Table \ref{table1}, a bilayer configuration allows for
ferroelectric control of spin polarization or AHE independently when it
is stacked with \(O = m_{001}\) and \(\mathbf{t}_{o}\) along the high
symmetry line GB or CA. We next perform density functional theory (DFT)
calculations on bilayer AgF\textsubscript{2}. We denote
B\textsubscript{1} and B\textsubscript{2} as bilayer configurations
which are stacked with
\({\widehat{O}}_{1} = \left\{ m_{001} \middle| (0.5,\  - 0.16) \right\}\)
and
\({\widehat{O}}_{2} = \left\{ m_{001} \middle| (0.5,\ 0.16) \right\}\),
respectively {[}see insets of Fig. \ref{Fig.2}(a) and Fig. S1
\cite{1013}{]}. Their interlayer magnetic coupling
includes two types: type-\(\text{I}\) and type-\(\text{\uppercase\expandafter{\romannumeral2}}\), as shown in
the insets of Figs. \ref{Fig.2}(b) and \ref{Fig.2}(c), respectively. DFT calculations show
that type-\(\text{I}\) is the magnetic ground state, with an energy 0.26
meV/f.u. lower than that of type-\(\text{\uppercase\expandafter{\romannumeral2}}\). The stacking breaks
inversion symmetry and transforms the layer group \(p2_{1}/b11\) of the
monolayer into \(pb11\) of the bilayer, inducing a polarization
\(P_{\ }\) in B\textsubscript{1} (B\textsubscript{2}) configuration
{[}see Fig. \ref{Fig.2}(a){]}. B\textsubscript{1} and B\textsubscript{2} are
connected by \(N^{-} = m_{001}\), enforcing exactly opposite \(P_{\ }\)
with the same amplitude (1.66 pC/m) for B\textsubscript{1} and
B\textsubscript{2} configurations. The energy barrier for the
ferroelectric switching from B\textsubscript{1} to B\textsubscript{2} is
17.1 meV/f.u.

In contrast to inversion symmetry breaking, stacking preserves the
\(\{ m_{100}|\frac{1}{2},\frac{1}{2}\}\) symmetry for B\textsubscript{1}
(B\textsubscript{2}) and thus inherits the altermagnetism, as shown in
Fig. \ref{Fig.2}(b). The maximal spin splitting is 51 meV for the two highest
occupied bands. B\textsubscript{1} and B\textsubscript{2} with the
type-\(\text{I}\) magnetic configuration are connected by
\(N_{s}^{-} = \{ 2_{\bot}||m_{001}\}\), which enforces them to have
opposite \(\mathbf{k}\)-dependent spin polarizations {[}Fig. \ref{Fig.2}(b){]}.
Therefore, the spin polarization can be switched by sliding
ferroelectricity in bilayer AgF\textsubscript{2}.

\begin{figure}
	\includegraphics[width=8.6cm]{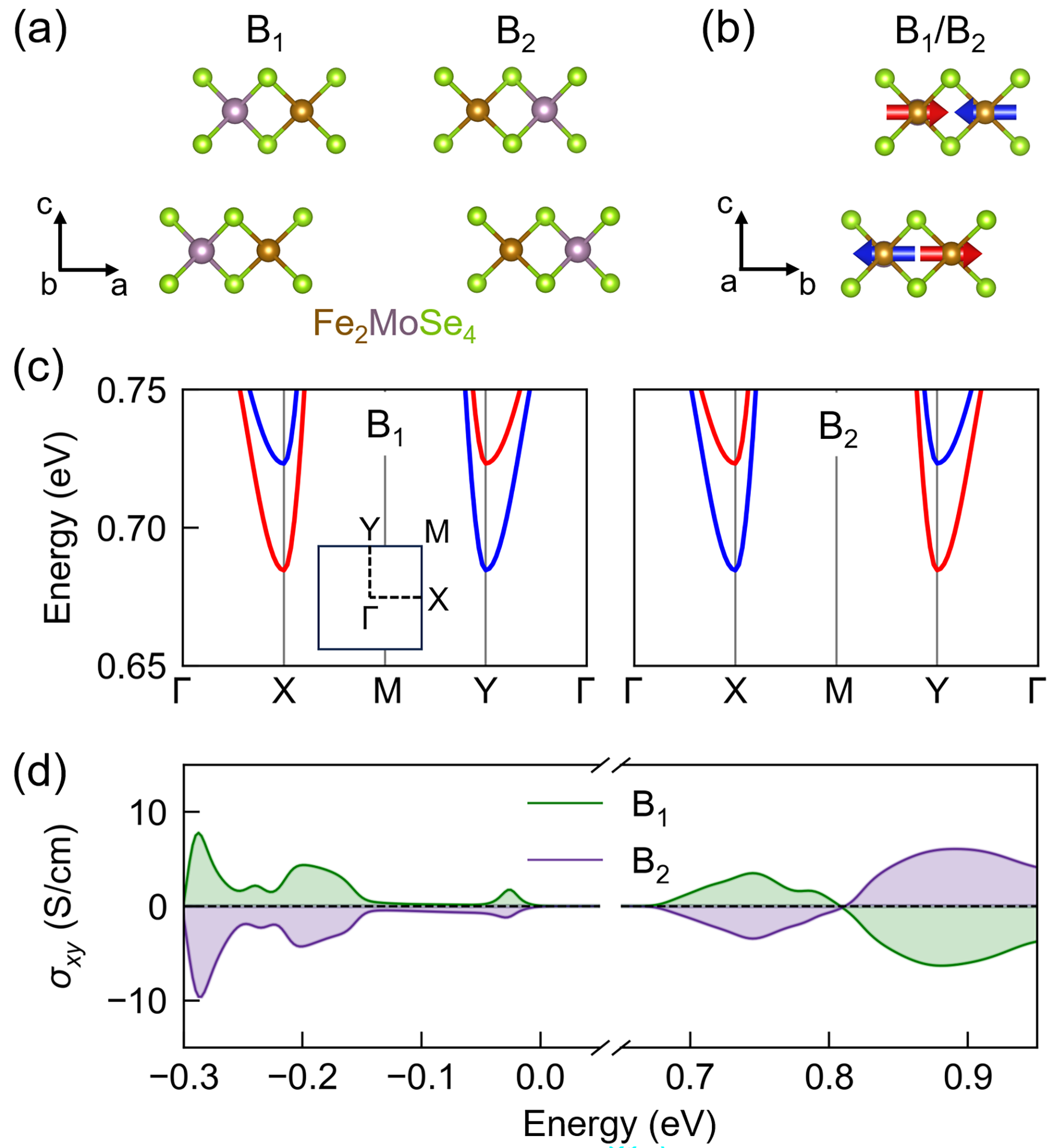}
	\caption{(a) The configurations of
Fe\textsubscript{2}MoSe\textsubscript{4} bilayers B\textsubscript{1} and
B\textsubscript{2}. (b) The same magnetic configurations of
B\textsubscript{1} and B\textsubscript{2}. The red and blue arrows
denote the magnetic moment along \(\lbrack 010\rbrack\) and
\(\lbrack 0\overline{1}0\rbrack\) directions, respectively. (c) The two
lowest unoccupied bands of B\textsubscript{1} and B\textsubscript{2},
calculated without considering SOC. The red and blue colors denote
opposite spins. The inset displays the Brillouin zone. (d) The anomalous
Hall conductivities of B\textsubscript{1} and B\textsubscript{2},
calculated with considering SOC. The valence band maximum is set to
zero.}
	\label{Fig.3}
\end{figure}

We next consider the SOC effect and discuss AHE in B\textsubscript{1}
and B\textsubscript{2}. By comparing the energies of B\textsubscript{1}
(B\textsubscript{2}) with the Néel vector aligned along {[}100{]},
{[}010{]} and {[}001{]} directions, we find that the easy axis is along
the {[}100{]} direction, as shown in the inset of Fig. \ref{Fig.2}(b).
B\textsubscript{1} (B\textsubscript{2}) has the magnetic point group
\(m' = \{ 1,m_{100}'\}\), which permits a nonzero anomalous Hall
conductivity \(\sigma_{xy}\). However, considering SOC,
B\textsubscript{1} and B\textsubscript{2} are connected by
\(N_{m}^{-} = m_{001}\), which does not switch the sign of Berry
curvature and thus cannot switch the AHE. Switching AHE requires
\(N_{m}^{-} = m_{001}'\), which necessitates tuning the magnetic
state of B\textsubscript{1} (B\textsubscript{2}) to the
type-\(\text{\uppercase\expandafter{\romannumeral2}}\) configuration, as shown in the inset of Fig. \ref{Fig.2}(c). We
take AgF\textsubscript{2} bilayers under a 4\% biaxial tensile strain as
an example, and denote B\textsubscript{1s} and B\textsubscript{2s} as
the bilayer B\textsubscript{1} and B\textsubscript{2} under strain,
respectively. For both B\textsubscript{1s} and B\textsubscript{2s}, the
type-\(\text{\uppercase\expandafter{\romannumeral2}}\) magnetic configuration is the ground state (see Table
S4 \cite{1013}). The easy axis remains along the {[}100{]} direction. While the sliding ferroelectricity is
preserved under strain (see Fig. S2 \cite{1013}).
B\textsubscript{1s} and B\textsubscript{2s} exhibit opposite anomalous
Hall conductivities as their connection operator
\(N_{m}^{-} = m_{001}'\) reverses the sign of Berry curvature {[}see
Fig. \ref{Fig.2}(c) and Fig. S2 \cite{1013}{]}. These results
indicate that the AHE can be controlled by the sliding ferroelectricity
in strained bilayer AgF\textsubscript{2}.

Similar to monolayer AgF\textsubscript{2}, monolayer
Fe\textsubscript{2}MoSe\textsubscript{4} is also an altermagnetic
candidate with the layer group \(p\overline{4}2m\) (No. 57)
\cite{1060}. However, unlike AgF\textsubscript{2}
bilayers connected by \(N^{-} = m_{001}\),
Fe\textsubscript{2}MoSe\textsubscript{4} bilayers allow for being
connected only by \(N^{-} = 2_{100}\) and thus enable spin polarization
and AHE to be simultaneously switched via sliding ferroelectricity (see
Table S2 \cite{1013}). We next perform DFT
calculations on bilayer Fe\textsubscript{2}MoSe\textsubscript{4}. As
shown in Fig. \ref{Fig.3}(a), we denote B\textsubscript{1} and B\textsubscript{2}
as the two bilayer Fe\textsubscript{2}MoSe\textsubscript{4}
configurations which are stacked with
\({\widehat{O}}_{1} = \left\{ 1 \middle| (\frac{1}{4},\frac{1}{4}) \right\}\)
and
\({\widehat{O}}_{2} = \left\{ 1 \middle| ( - \frac{1}{4},\frac{1}{4}) \right\}\),
respectively. The stacking transforms the layer group
\(p\overline{4}2m\) of the monolayer into \(pm11\) of the bilayer,
inducing a polarization \(P_{\ }\) in B\textsubscript{1}
(B\textsubscript{2}) configuration. B\textsubscript{1} and
B\textsubscript{2} are connected by \(N^{-} = 2_{100}\), enforcing
exactly opposite \(P_{\ }\) with the same amplitude (0.14 pC/m) for
them.

Fig. \ref{Fig.3}(b) shows the same magnetic configuration for B\textsubscript{1}
and B\textsubscript{2}. In this magnetic configuration, they are
connected by \(N_{s}^{-} = \{ 2_{\bot}||2_{100}\}\) without considering
SOC, and by \(N_{m}^{-} = 2_{100}\) when SOC is included. As shown in
Fig. \ref{Fig.3}(c), \(N_{s}^{-}\) enforces that B\textsubscript{1} and
B\textsubscript{2} exhibit opposite \(\mathbf{k}\)-dependent spin
splitting with the maximal value of 39 meV for the two lowest unoccupied
bands. When SOC is included, B\textsubscript{1} and B\textsubscript{2}
exhibit opposite anomalous Hall conductivities enforced by
\(N_{m}^{-} = 2_{100}\), as shown in Fig. \ref{Fig.3}(d). Therefore, unlike
bilayer AgF\textsubscript{2}, both spin polarization and AHE of bilayer
Fe\textsubscript{2}MoSe\textsubscript{4} can be simultaneously switched
via ferroelectric switching. Given that AHE is more accessible to detect
experimentally than spin polarization \cite{845},  the
simultaneous control of spin polarization and AHE enables spin
polarization switching to be more observable.

\emph{Generalized stacking rule for compensated ferrimagnets}---We now generalize the stacking rule to collinear compensated ferrimagnet
(CFiM) \cite{850, 495, 1052, 873}. Unlike
altermagnets, where zero net magnetization is enforced by symmetry,
collinear CFiMs exhibit zero net magnetization due to appropriate
electron filling in semiconducting or half-metallic states. We focus on
collinear CFiM bilayers, which consist of two ferromagnetic monolayers
with opposite spins. Such CFiM are permitted for all stacking orders
that support sliding ferroelectricity, as listed in Tables S1-S3
\cite{1013}. Interestingly, in ferroelectric CFiM
bilayers with out-of-plane magnetic configurations, both spin
polarization and AHE can be simultaneously controlled via sliding
ferroelectricity (see Appendix). Candidate materials include
experimentally synthesized monolayer CrI\textsubscript{3}
\cite{76},  CrGeTe\textsubscript{3}
\cite{445},  MnBi\textsubscript{2}Te\textsubscript{4}
\cite{1069, 1067},  etc. Taking CrI\textsubscript{3} as
an example, DFT calculations demonstrate that spin polarization and AHE
in CFiM bilayer CrI\textsubscript{3} can be simultaneously switched via
sliding ferroelectricity (see details in Sec. 4
\cite{1013}).

\emph{Summary and discussion}---Recent studies have proposed electric control of spin polarization in
altermagnets \cite{706, 1055, 1056}. Ref.
\cite{706} introduces gate-field control of the spin
degree of freedom in altermagnets based on spin-valley-layer coupling.
The spin control relies on inducing opposite valley polarization, which
requires a constant electric field and is thus volatile. Ref.
\cite{1055} has proposed controlling the existence of
altermagnetism through the phase transition between antiferroelectric
and ferroelectric states. Ref. \cite{1056} has
proposed ferroelectric control of the sign of spin polarization in
altermagnets. Compared with these studies, our work extends the range of
electric controllable functionalities from spin polarization to AHE,
and, furthermore, the simultaneous control of both spin polarization and
AHE. Moreover, we generalize the stacking rule beyond altermagnets to
include another type of unconventional magnets, i.e., collinear CFiMs.

In fact, the general rule is applicable not only to collinear
unconventional magnets (such as altermagnets and collinear CFiMs), but
also to noncollinear and noncoplanar ones. Additionally, the rule can be
extended beyond spin polarization and AHE to other spin-dependent
properties (e.g., magneto-optical effects) that can be switched by the
connection operator. Our work provides a general strategy for designing
unconventional multiferroic bilayer systems and developing
energy-efficient antiferromagnetic spintronic devices.

\begin{acknowledgments}
This work is supported by the National Key Research and Development
Program of China (2022YFA1204100), the National Natural Science
Foundation of China (62488201 and 12274194), Shenzhen Science and
Technology Program (RCJC20221008092722009 and No. 20231117091158001),
the Innovative Team of General Higher Educational Institutes in
Guangdong Province (2020KCXTD001), and Guangdong Provincial Quantum
Science Strategic Initiative (GDZX2401002).
\end{acknowledgments}


\begin{thebibliography}{96}%
\makeatletter
\providecommand \@ifxundefined [1]{%
 \@ifx{#1\undefined}
}%
\providecommand \@ifnum [1]{%
 \ifnum #1\expandafter \@firstoftwo
 \else \expandafter \@secondoftwo
 \fi
}%
\providecommand \@ifx [1]{%
 \ifx #1\expandafter \@firstoftwo
 \else \expandafter \@secondoftwo
 \fi
}%
\providecommand \natexlab [1]{#1}%
\providecommand \enquote  [1]{``#1''}%
\providecommand \bibnamefont  [1]{#1}%
\providecommand \bibfnamefont [1]{#1}%
\providecommand \citenamefont [1]{#1}%
\providecommand \href@noop [0]{\@secondoftwo}%
\providecommand \href [0]{\begingroup \@sanitize@url \@href}%
\providecommand \@href[1]{\@@startlink{#1}\@@href}%
\providecommand \@@href[1]{\endgroup#1\@@endlink}%
\providecommand \@sanitize@url [0]{\catcode `\\12\catcode `\$12\catcode `\&12\catcode `\#12\catcode `\^12\catcode `\_12\catcode `\%12\relax}%
\providecommand \@@startlink[1]{}%
\providecommand \@@endlink[0]{}%
\providecommand \url  [0]{\begingroup\@sanitize@url \@url }%
\providecommand \@url [1]{\endgroup\@href {#1}{\urlprefix }}%
\providecommand \urlprefix  [0]{URL }%
\providecommand \Eprint [0]{\href }%
\providecommand \doibase [0]{https://doi.org/}%
\providecommand \selectlanguage [0]{\@gobble}%
\providecommand \bibinfo  [0]{\@secondoftwo}%
\providecommand \bibfield  [0]{\@secondoftwo}%
\providecommand \translation [1]{[#1]}%
\providecommand \BibitemOpen [0]{}%
\providecommand \bibitemStop [0]{}%
\providecommand \bibitemNoStop [0]{.\EOS\space}%
\providecommand \EOS [0]{\spacefactor3000\relax}%
\providecommand \BibitemShut  [1]{\csname bibitem#1\endcsname}%
\let\auto@bib@innerbib\@empty
\bibitem [{\citenamefont {Liu}\ \emph {et~al.}(2025{\natexlab{a}})\citenamefont {Liu}, \citenamefont {Dai},\ and\ \citenamefont {Blügel}}]{1000}%
  \BibitemOpen
  \bibfield  {author} {\bibinfo {author} {\bibfnamefont {Q.}~\bibnamefont {Liu}}, \bibinfo {author} {\bibfnamefont {X.}~\bibnamefont {Dai}},\ and\ \bibinfo {author} {\bibfnamefont {S.}~\bibnamefont {Blügel}},\ }\href@noop {} {\bibfield  {journal} {\bibinfo  {journal} {Nat. Phys.}\ }\textbf {\bibinfo {volume} {21}},\ \bibinfo {pages} {329} (\bibinfo {year} {2025}{\natexlab{a}})}\BibitemShut {NoStop}%
\bibitem [{\citenamefont {Chen}\ \emph {et~al.}(2025)\citenamefont {Chen}, \citenamefont {Liu}, \citenamefont {Liu}, \citenamefont {Yu}, \citenamefont {Ren}, \citenamefont {Li}, \citenamefont {Zhang},\ and\ \citenamefont {Liu}}]{1042}%
  \BibitemOpen
  \bibfield  {author} {\bibinfo {author} {\bibfnamefont {X.}~\bibnamefont {Chen}}, \bibinfo {author} {\bibfnamefont {Y.}~\bibnamefont {Liu}}, \bibinfo {author} {\bibfnamefont {P.}~\bibnamefont {Liu}}, \bibinfo {author} {\bibfnamefont {Y.}~\bibnamefont {Yu}}, \bibinfo {author} {\bibfnamefont {J.}~\bibnamefont {Ren}}, \bibinfo {author} {\bibfnamefont {J.}~\bibnamefont {Li}}, \bibinfo {author} {\bibfnamefont {A.}~\bibnamefont {Zhang}},\ and\ \bibinfo {author} {\bibfnamefont {Q.}~\bibnamefont {Liu}},\ }\href@noop {} {\bibfield  {journal} {\bibinfo  {journal} {Nature}\ }\textbf {\bibinfo {volume} {640}},\ \bibinfo {pages} {349–354} (\bibinfo {year} {2025})}\BibitemShut {NoStop}%
\bibitem [{\citenamefont {Chen}\ \emph {et~al.}(2024)\citenamefont {Chen}, \citenamefont {Ren}, \citenamefont {Zhu}, \citenamefont {Yu}, \citenamefont {Zhang}, \citenamefont {Liu}, \citenamefont {Li}, \citenamefont {Liu}, \citenamefont {Li},\ and\ \citenamefont {Liu}}]{743}%
  \BibitemOpen
  \bibfield  {author} {\bibinfo {author} {\bibfnamefont {X.}~\bibnamefont {Chen}}, \bibinfo {author} {\bibfnamefont {J.}~\bibnamefont {Ren}}, \bibinfo {author} {\bibfnamefont {Y.}~\bibnamefont {Zhu}}, \bibinfo {author} {\bibfnamefont {Y.}~\bibnamefont {Yu}}, \bibinfo {author} {\bibfnamefont {A.}~\bibnamefont {Zhang}}, \bibinfo {author} {\bibfnamefont {P.}~\bibnamefont {Liu}}, \bibinfo {author} {\bibfnamefont {J.}~\bibnamefont {Li}}, \bibinfo {author} {\bibfnamefont {Y.}~\bibnamefont {Liu}}, \bibinfo {author} {\bibfnamefont {C.}~\bibnamefont {Li}},\ and\ \bibinfo {author} {\bibfnamefont {Q.}~\bibnamefont {Liu}},\ }\href@noop {} {\bibfield  {journal} {\bibinfo  {journal} {Phys. Rev. X}\ }\textbf {\bibinfo {volume} {14}},\ \bibinfo {pages} {031038} (\bibinfo {year} {2024})}\BibitemShut {NoStop}%
\bibitem [{\citenamefont {Jiang}\ \emph {et~al.}(2024)\citenamefont {Jiang}, \citenamefont {Song}, \citenamefont {Zhu}, \citenamefont {Fang}, \citenamefont {Weng}, \citenamefont {Liu}, \citenamefont {Yang},\ and\ \citenamefont {Fang}}]{773}%
  \BibitemOpen
  \bibfield  {author} {\bibinfo {author} {\bibfnamefont {Y.}~\bibnamefont {Jiang}}, \bibinfo {author} {\bibfnamefont {Z.}~\bibnamefont {Song}}, \bibinfo {author} {\bibfnamefont {T.}~\bibnamefont {Zhu}}, \bibinfo {author} {\bibfnamefont {Z.}~\bibnamefont {Fang}}, \bibinfo {author} {\bibfnamefont {H.}~\bibnamefont {Weng}}, \bibinfo {author} {\bibfnamefont {Z.-X.}\ \bibnamefont {Liu}}, \bibinfo {author} {\bibfnamefont {J.}~\bibnamefont {Yang}},\ and\ \bibinfo {author} {\bibfnamefont {C.}~\bibnamefont {Fang}},\ }\href@noop {} {\bibfield  {journal} {\bibinfo  {journal} {Phys. Rev. X}\ }\textbf {\bibinfo {volume} {14}},\ \bibinfo {pages} {031039} (\bibinfo {year} {2024})}\BibitemShut {NoStop}%
\bibitem [{\citenamefont {Krempaský}\ \emph {et~al.}(2024)\citenamefont {Krempaský}, \citenamefont {Šmejkal}, \citenamefont {D’Souza}, \citenamefont {Hajlaoui}, \citenamefont {Springholz}, \citenamefont {Uhlířová}, \citenamefont {Alarab}, \citenamefont {Constantinou}, \citenamefont {Strocov}, \citenamefont {Usanov}, \citenamefont {Pudelko}, \citenamefont {González-Hernández}, \citenamefont {Birk~Hellenes}, \citenamefont {Jansa}, \citenamefont {Reichlová}, \citenamefont {Šobáň}, \citenamefont {Gonzalez~Betancourt}, \citenamefont {Wadley}, \citenamefont {Sinova}, \citenamefont {Kriegner}, \citenamefont {Minár}, \citenamefont {Dil},\ and\ \citenamefont {Jungwirth}}]{669}%
  \BibitemOpen
  \bibfield  {author} {\bibinfo {author} {\bibfnamefont {J.}~\bibnamefont {Krempaský}}, \bibinfo {author} {\bibfnamefont {L.}~\bibnamefont {Šmejkal}}, \bibinfo {author} {\bibfnamefont {S.~W.}\ \bibnamefont {D’Souza}}, \bibinfo {author} {\bibfnamefont {M.}~\bibnamefont {Hajlaoui}}, \bibinfo {author} {\bibfnamefont {G.}~\bibnamefont {Springholz}}, \bibinfo {author} {\bibfnamefont {K.}~\bibnamefont {Uhlířová}}, \bibinfo {author} {\bibfnamefont {F.}~\bibnamefont {Alarab}}, \bibinfo {author} {\bibfnamefont {P.~C.}\ \bibnamefont {Constantinou}}, \bibinfo {author} {\bibfnamefont {V.}~\bibnamefont {Strocov}}, \bibinfo {author} {\bibfnamefont {D.}~\bibnamefont {Usanov}}, \bibinfo {author} {\bibfnamefont {W.~R.}\ \bibnamefont {Pudelko}}, \bibinfo {author} {\bibfnamefont {R.}~\bibnamefont {González-Hernández}}, \bibinfo {author} {\bibfnamefont {A.}~\bibnamefont {Birk~Hellenes}}, \bibinfo {author} {\bibfnamefont {Z.}~\bibnamefont {Jansa}}, \bibinfo {author} {\bibfnamefont {H.}~\bibnamefont {Reichlová}}, \bibinfo {author} {\bibfnamefont {Z.}~\bibnamefont {Šobáň}}, \bibinfo {author} {\bibfnamefont {R.~D.}\ \bibnamefont {Gonzalez~Betancourt}}, \bibinfo {author} {\bibfnamefont {P.}~\bibnamefont {Wadley}}, \bibinfo {author} {\bibfnamefont {J.}~\bibnamefont {Sinova}}, \bibinfo {author} {\bibfnamefont {D.}~\bibnamefont {Kriegner}}, \bibinfo {author} {\bibfnamefont {J.}~\bibnamefont {Minár}}, \bibinfo {author} {\bibfnamefont {J.~H.}\ \bibnamefont {Dil}},\ and\ \bibinfo {author} {\bibfnamefont {T.}~\bibnamefont {Jungwirth}},\ }\href@noop {} {\bibfield  {journal} {\bibinfo  {journal} {Nature}\ }\textbf {\bibinfo {volume} {626}},\ \bibinfo {pages} {517} (\bibinfo {year} {2024})}\BibitemShut {NoStop}%
\bibitem [{\citenamefont {Liu}\ \emph {et~al.}(2022)\citenamefont {Liu}, \citenamefont {Li}, \citenamefont {Han}, \citenamefont {Wan},\ and\ \citenamefont {Liu}}]{147}%
  \BibitemOpen
  \bibfield  {author} {\bibinfo {author} {\bibfnamefont {P.}~\bibnamefont {Liu}}, \bibinfo {author} {\bibfnamefont {J.}~\bibnamefont {Li}}, \bibinfo {author} {\bibfnamefont {J.}~\bibnamefont {Han}}, \bibinfo {author} {\bibfnamefont {X.}~\bibnamefont {Wan}},\ and\ \bibinfo {author} {\bibfnamefont {Q.}~\bibnamefont {Liu}},\ }\href@noop {} {\bibfield  {journal} {\bibinfo  {journal} {Phys. Rev. X}\ }\textbf {\bibinfo {volume} {12}},\ \bibinfo {pages} {021016} (\bibinfo {year} {2022})}\BibitemShut {NoStop}%
\bibitem [{\citenamefont {Šmejkal}\ \emph {et~al.}(2022{\natexlab{a}})\citenamefont {Šmejkal}, \citenamefont {Sinova},\ and\ \citenamefont {Jungwirth}}]{580}%
  \BibitemOpen
  \bibfield  {author} {\bibinfo {author} {\bibfnamefont {L.}~\bibnamefont {Šmejkal}}, \bibinfo {author} {\bibfnamefont {J.}~\bibnamefont {Sinova}},\ and\ \bibinfo {author} {\bibfnamefont {T.}~\bibnamefont {Jungwirth}},\ }\href@noop {} {\bibfield  {journal} {\bibinfo  {journal} {Phys. Rev. X}\ }\textbf {\bibinfo {volume} {12}},\ \bibinfo {pages} {031042} (\bibinfo {year} {2022}{\natexlab{a}})}\BibitemShut {NoStop}%
\bibitem [{\citenamefont {Šmejkal}\ \emph {et~al.}(2022{\natexlab{b}})\citenamefont {Šmejkal}, \citenamefont {Sinova},\ and\ \citenamefont {Jungwirth}}]{255}%
  \BibitemOpen
  \bibfield  {author} {\bibinfo {author} {\bibfnamefont {L.}~\bibnamefont {Šmejkal}}, \bibinfo {author} {\bibfnamefont {J.}~\bibnamefont {Sinova}},\ and\ \bibinfo {author} {\bibfnamefont {T.}~\bibnamefont {Jungwirth}},\ }\href@noop {} {\bibfield  {journal} {\bibinfo  {journal} {Phys. Rev. X}\ }\textbf {\bibinfo {volume} {12}},\ \bibinfo {pages} {040501} (\bibinfo {year} {2022}{\natexlab{b}})}\BibitemShut {NoStop}%
\bibitem [{\citenamefont {Xiao}\ \emph {et~al.}(2024)\citenamefont {Xiao}, \citenamefont {Zhao}, \citenamefont {Li}, \citenamefont {Shindou},\ and\ \citenamefont {Song}}]{898}%
  \BibitemOpen
  \bibfield  {author} {\bibinfo {author} {\bibfnamefont {Z.}~\bibnamefont {Xiao}}, \bibinfo {author} {\bibfnamefont {J.}~\bibnamefont {Zhao}}, \bibinfo {author} {\bibfnamefont {Y.}~\bibnamefont {Li}}, \bibinfo {author} {\bibfnamefont {R.}~\bibnamefont {Shindou}},\ and\ \bibinfo {author} {\bibfnamefont {Z.-D.}\ \bibnamefont {Song}},\ }\href@noop {} {\bibfield  {journal} {\bibinfo  {journal} {Phys. Rev. X}\ }\textbf {\bibinfo {volume} {14}},\ \bibinfo {pages} {031037} (\bibinfo {year} {2024})}\BibitemShut {NoStop}%
\bibitem [{\citenamefont {Zhu}\ \emph {et~al.}(2024)\citenamefont {Zhu}, \citenamefont {Chen}, \citenamefont {Liu}, \citenamefont {Liu}, \citenamefont {Liu}, \citenamefont {Zha}, \citenamefont {Qu}, \citenamefont {Hong}, \citenamefont {Li}, \citenamefont {Jiang}, \citenamefont {Ma}, \citenamefont {Hao}, \citenamefont {Zhu}, \citenamefont {Liu}, \citenamefont {Zeng}, \citenamefont {Jayaram}, \citenamefont {Lenger}, \citenamefont {Ding}, \citenamefont {Mo}, \citenamefont {Tanaka}, \citenamefont {Arita}, \citenamefont {Liu}, \citenamefont {Ye}, \citenamefont {Shen}, \citenamefont {Wrachtrup}, \citenamefont {Huang}, \citenamefont {He}, \citenamefont {Qiao}, \citenamefont {Liu},\ and\ \citenamefont {Liu}}]{671}%
  \BibitemOpen
  \bibfield  {author} {\bibinfo {author} {\bibfnamefont {Y.-P.}\ \bibnamefont {Zhu}}, \bibinfo {author} {\bibfnamefont {X.}~\bibnamefont {Chen}}, \bibinfo {author} {\bibfnamefont {X.-R.}\ \bibnamefont {Liu}}, \bibinfo {author} {\bibfnamefont {Y.}~\bibnamefont {Liu}}, \bibinfo {author} {\bibfnamefont {P.}~\bibnamefont {Liu}}, \bibinfo {author} {\bibfnamefont {H.}~\bibnamefont {Zha}}, \bibinfo {author} {\bibfnamefont {G.}~\bibnamefont {Qu}}, \bibinfo {author} {\bibfnamefont {C.}~\bibnamefont {Hong}}, \bibinfo {author} {\bibfnamefont {J.}~\bibnamefont {Li}}, \bibinfo {author} {\bibfnamefont {Z.}~\bibnamefont {Jiang}}, \bibinfo {author} {\bibfnamefont {X.-M.}\ \bibnamefont {Ma}}, \bibinfo {author} {\bibfnamefont {Y.-J.}\ \bibnamefont {Hao}}, \bibinfo {author} {\bibfnamefont {M.-Y.}\ \bibnamefont {Zhu}}, \bibinfo {author} {\bibfnamefont {W.}~\bibnamefont {Liu}}, \bibinfo {author} {\bibfnamefont {M.}~\bibnamefont {Zeng}}, \bibinfo {author} {\bibfnamefont {S.}~\bibnamefont {Jayaram}}, \bibinfo {author} {\bibfnamefont {M.}~\bibnamefont {Lenger}}, \bibinfo {author} {\bibfnamefont {J.}~\bibnamefont {Ding}}, \bibinfo {author} {\bibfnamefont {S.}~\bibnamefont {Mo}}, \bibinfo {author} {\bibfnamefont {K.}~\bibnamefont {Tanaka}}, \bibinfo {author} {\bibfnamefont {M.}~\bibnamefont {Arita}}, \bibinfo {author} {\bibfnamefont {Z.}~\bibnamefont {Liu}}, \bibinfo {author} {\bibfnamefont {M.}~\bibnamefont {Ye}}, \bibinfo {author} {\bibfnamefont {D.}~\bibnamefont {Shen}}, \bibinfo {author} {\bibfnamefont {J.}~\bibnamefont {Wrachtrup}}, \bibinfo {author} {\bibfnamefont {Y.}~\bibnamefont {Huang}}, \bibinfo {author} {\bibfnamefont {R.-H.}\ \bibnamefont {He}}, \bibinfo {author} {\bibfnamefont {S.}~\bibnamefont {Qiao}}, \bibinfo {author} {\bibfnamefont {Q.}~\bibnamefont {Liu}},\ and\ \bibinfo {author} {\bibfnamefont {C.}~\bibnamefont {Liu}},\ }\href@noop {} {\bibfield  {journal} {\bibinfo  {journal} {Nature}\ }\textbf {\bibinfo {volume} {626}},\ \bibinfo {pages} {523} (\bibinfo {year} {2024})}\BibitemShut {NoStop}%
\bibitem [{\citenamefont {Hayami}\ \emph {et~al.}(2019)\citenamefont {Hayami}, \citenamefont {Yanagi},\ and\ \citenamefont {Kusunose}}]{897}%
  \BibitemOpen
  \bibfield  {author} {\bibinfo {author} {\bibfnamefont {S.}~\bibnamefont {Hayami}}, \bibinfo {author} {\bibfnamefont {Y.}~\bibnamefont {Yanagi}},\ and\ \bibinfo {author} {\bibfnamefont {H.}~\bibnamefont {Kusunose}},\ }\href@noop {} {\bibfield  {journal} {\bibinfo  {journal} {J. Phys. Soc. Jpn.}\ }\textbf {\bibinfo {volume} {88}},\ \bibinfo {pages} {123702} (\bibinfo {year} {2019})}\BibitemShut {NoStop}%
\bibitem [{\citenamefont {Ma}\ \emph {et~al.}(2021)\citenamefont {Ma}, \citenamefont {Hu}, \citenamefont {Li}, \citenamefont {Liu}, \citenamefont {Yao}, \citenamefont {Jia},\ and\ \citenamefont {Liu}}]{658}%
  \BibitemOpen
  \bibfield  {author} {\bibinfo {author} {\bibfnamefont {H.-Y.}\ \bibnamefont {Ma}}, \bibinfo {author} {\bibfnamefont {M.}~\bibnamefont {Hu}}, \bibinfo {author} {\bibfnamefont {N.}~\bibnamefont {Li}}, \bibinfo {author} {\bibfnamefont {J.}~\bibnamefont {Liu}}, \bibinfo {author} {\bibfnamefont {W.}~\bibnamefont {Yao}}, \bibinfo {author} {\bibfnamefont {J.-F.}\ \bibnamefont {Jia}},\ and\ \bibinfo {author} {\bibfnamefont {J.}~\bibnamefont {Liu}},\ }\href@noop {} {\bibfield  {journal} {\bibinfo  {journal} {Nat. Commun.}\ }\textbf {\bibinfo {volume} {12}},\ \bibinfo {pages} {2846} (\bibinfo {year} {2021})}\BibitemShut {NoStop}%
\bibitem [{\citenamefont {Mazin}\ \emph {et~al.}(2021)\citenamefont {Mazin}, \citenamefont {Koepernik}, \citenamefont {Johannes}, \citenamefont {Gonzalez-Hernandez},\ and\ \citenamefont {Smejkal}}]{894}%
  \BibitemOpen
  \bibfield  {author} {\bibinfo {author} {\bibfnamefont {I.}~\bibnamefont {Mazin}}, \bibinfo {author} {\bibfnamefont {K.}~\bibnamefont {Koepernik}}, \bibinfo {author} {\bibfnamefont {M.~D.}\ \bibnamefont {Johannes}}, \bibinfo {author} {\bibfnamefont {R.}~\bibnamefont {Gonzalez-Hernandez}},\ and\ \bibinfo {author} {\bibfnamefont {L.}~\bibnamefont {Smejkal}},\ }\href@noop {} {\bibfield  {journal} {\bibinfo  {journal} {Proc Natl Acad Sci U S A}\ }\textbf {\bibinfo {volume} {118}},\ \bibinfo {pages} {e2108924118} (\bibinfo {year} {2021})}\BibitemShut {NoStop}%
\bibitem [{\citenamefont {Šmejkal}\ \emph {et~al.}(2020)\citenamefont {Šmejkal}, \citenamefont {González-Hernández}, \citenamefont {Jungwirth},\ and\ \citenamefont {Sinova}}]{848}%
  \BibitemOpen
  \bibfield  {author} {\bibinfo {author} {\bibfnamefont {L.}~\bibnamefont {Šmejkal}}, \bibinfo {author} {\bibfnamefont {R.}~\bibnamefont {González-Hernández}}, \bibinfo {author} {\bibfnamefont {T.}~\bibnamefont {Jungwirth}},\ and\ \bibinfo {author} {\bibfnamefont {J.}~\bibnamefont {Sinova}},\ }\href@noop {} {\bibfield  {journal} {\bibinfo  {journal} {Sci. Adv.}\ }\textbf {\bibinfo {volume} {6}},\ \bibinfo {pages} {eaaz8809} (\bibinfo {year} {2020})}\BibitemShut {NoStop}%
\bibitem [{\citenamefont {Yuan}\ \emph {et~al.}(2020)\citenamefont {Yuan}, \citenamefont {Wang}, \citenamefont {Luo}, \citenamefont {Rashba},\ and\ \citenamefont {Zunger}}]{693}%
  \BibitemOpen
  \bibfield  {author} {\bibinfo {author} {\bibfnamefont {L.-D.}\ \bibnamefont {Yuan}}, \bibinfo {author} {\bibfnamefont {Z.}~\bibnamefont {Wang}}, \bibinfo {author} {\bibfnamefont {J.-W.}\ \bibnamefont {Luo}}, \bibinfo {author} {\bibfnamefont {E.~I.}\ \bibnamefont {Rashba}},\ and\ \bibinfo {author} {\bibfnamefont {A.}~\bibnamefont {Zunger}},\ }\href@noop {} {\bibfield  {journal} {\bibinfo  {journal} {Phys. Rev. B}\ }\textbf {\bibinfo {volume} {102}},\ \bibinfo {pages} {014422} (\bibinfo {year} {2020})}\BibitemShut {NoStop}%
\bibitem [{\citenamefont {Bai}\ \emph {et~al.}(2022)\citenamefont {Bai}, \citenamefont {Han}, \citenamefont {Feng}, \citenamefont {Zhou}, \citenamefont {Su}, \citenamefont {Wang}, \citenamefont {Liao}, \citenamefont {Zhu}, \citenamefont {Chen}, \citenamefont {Pan}, \citenamefont {Fan},\ and\ \citenamefont {Song}}]{983}%
  \BibitemOpen
  \bibfield  {author} {\bibinfo {author} {\bibfnamefont {H.}~\bibnamefont {Bai}}, \bibinfo {author} {\bibfnamefont {L.}~\bibnamefont {Han}}, \bibinfo {author} {\bibfnamefont {X.~Y.}\ \bibnamefont {Feng}}, \bibinfo {author} {\bibfnamefont {Y.~J.}\ \bibnamefont {Zhou}}, \bibinfo {author} {\bibfnamefont {R.~X.}\ \bibnamefont {Su}}, \bibinfo {author} {\bibfnamefont {Q.}~\bibnamefont {Wang}}, \bibinfo {author} {\bibfnamefont {L.~Y.}\ \bibnamefont {Liao}}, \bibinfo {author} {\bibfnamefont {W.~X.}\ \bibnamefont {Zhu}}, \bibinfo {author} {\bibfnamefont {X.~Z.}\ \bibnamefont {Chen}}, \bibinfo {author} {\bibfnamefont {F.}~\bibnamefont {Pan}}, \bibinfo {author} {\bibfnamefont {X.~L.}\ \bibnamefont {Fan}},\ and\ \bibinfo {author} {\bibfnamefont {C.}~\bibnamefont {Song}},\ }\href@noop {} {\bibfield  {journal} {\bibinfo  {journal} {Phys. Rev. Lett.}\ }\textbf {\bibinfo {volume} {128}},\ \bibinfo {pages} {197202} (\bibinfo {year} {2022})}\BibitemShut {NoStop}%
\bibitem [{\citenamefont {Bai}\ \emph {et~al.}(2024)\citenamefont {Bai}, \citenamefont {Feng}, \citenamefont {Liu}, \citenamefont {Smejkal}, \citenamefont {Mokrousov},\ and\ \citenamefont {Yao}}]{793}%
  \BibitemOpen
  \bibfield  {author} {\bibinfo {author} {\bibfnamefont {L.}~\bibnamefont {Bai}}, \bibinfo {author} {\bibfnamefont {W.~X.}\ \bibnamefont {Feng}}, \bibinfo {author} {\bibfnamefont {S.~Y.}\ \bibnamefont {Liu}}, \bibinfo {author} {\bibfnamefont {L.}~\bibnamefont {Smejkal}}, \bibinfo {author} {\bibfnamefont {Y.}~\bibnamefont {Mokrousov}},\ and\ \bibinfo {author} {\bibfnamefont {Y.~G.}\ \bibnamefont {Yao}},\ }\href@noop {} {\bibfield  {journal} {\bibinfo  {journal} {Adv. Funct. Mater.}\ }\textbf {\bibinfo {volume} {34}},\ \bibinfo {pages} {2409327} (\bibinfo {year} {2024})}\BibitemShut {NoStop}%
\bibitem [{\citenamefont {Gonzalez-Hernandez}\ \emph {et~al.}(2021)\citenamefont {Gonzalez-Hernandez}, \citenamefont {Smejkal}, \citenamefont {Vyborny}, \citenamefont {Yahagi}, \citenamefont {Sinova}, \citenamefont {Jungwirth},\ and\ \citenamefont {Zelezny}}]{986}%
  \BibitemOpen
  \bibfield  {author} {\bibinfo {author} {\bibfnamefont {R.}~\bibnamefont {Gonzalez-Hernandez}}, \bibinfo {author} {\bibfnamefont {L.}~\bibnamefont {Smejkal}}, \bibinfo {author} {\bibfnamefont {K.}~\bibnamefont {Vyborny}}, \bibinfo {author} {\bibfnamefont {Y.}~\bibnamefont {Yahagi}}, \bibinfo {author} {\bibfnamefont {J.}~\bibnamefont {Sinova}}, \bibinfo {author} {\bibfnamefont {T.}~\bibnamefont {Jungwirth}},\ and\ \bibinfo {author} {\bibfnamefont {J.}~\bibnamefont {Zelezny}},\ }\href@noop {} {\bibfield  {journal} {\bibinfo  {journal} {Phys. Rev. Lett.}\ }\textbf {\bibinfo {volume} {126}},\ \bibinfo {pages} {127701} (\bibinfo {year} {2021})}\BibitemShut {NoStop}%
\bibitem [{\citenamefont {Naka}\ \emph {et~al.}(2019)\citenamefont {Naka}, \citenamefont {Hayami}, \citenamefont {Kusunose}, \citenamefont {Yanagi}, \citenamefont {Motome},\ and\ \citenamefont {Seo}}]{870}%
  \BibitemOpen
  \bibfield  {author} {\bibinfo {author} {\bibfnamefont {M.}~\bibnamefont {Naka}}, \bibinfo {author} {\bibfnamefont {S.}~\bibnamefont {Hayami}}, \bibinfo {author} {\bibfnamefont {H.}~\bibnamefont {Kusunose}}, \bibinfo {author} {\bibfnamefont {Y.}~\bibnamefont {Yanagi}}, \bibinfo {author} {\bibfnamefont {Y.}~\bibnamefont {Motome}},\ and\ \bibinfo {author} {\bibfnamefont {H.}~\bibnamefont {Seo}},\ }\href@noop {} {\bibfield  {journal} {\bibinfo  {journal} {Nat. Commun.}\ }\textbf {\bibinfo {volume} {10}},\ \bibinfo {pages} {4305} (\bibinfo {year} {2019})}\BibitemShut {NoStop}%
\bibitem [{\citenamefont {Naka}\ \emph {et~al.}(2021)\citenamefont {Naka}, \citenamefont {Motome},\ and\ \citenamefont {Seo}}]{987}%
  \BibitemOpen
  \bibfield  {author} {\bibinfo {author} {\bibfnamefont {M.}~\bibnamefont {Naka}}, \bibinfo {author} {\bibfnamefont {Y.}~\bibnamefont {Motome}},\ and\ \bibinfo {author} {\bibfnamefont {H.}~\bibnamefont {Seo}},\ }\href@noop {} {\bibfield  {journal} {\bibinfo  {journal} {Phys. Rev. B}\ }\textbf {\bibinfo {volume} {103}},\ \bibinfo {pages} {125114} (\bibinfo {year} {2021})}\BibitemShut {NoStop}%
\bibitem [{\citenamefont {Shao}\ \emph {et~al.}(2021{\natexlab{a}})\citenamefont {Shao}, \citenamefont {Zhang}, \citenamefont {Li}, \citenamefont {Eom},\ and\ \citenamefont {Tsymbal}}]{957}%
  \BibitemOpen
  \bibfield  {author} {\bibinfo {author} {\bibfnamefont {D.~F.}\ \bibnamefont {Shao}}, \bibinfo {author} {\bibfnamefont {S.~H.}\ \bibnamefont {Zhang}}, \bibinfo {author} {\bibfnamefont {M.}~\bibnamefont {Li}}, \bibinfo {author} {\bibfnamefont {C.~B.}\ \bibnamefont {Eom}},\ and\ \bibinfo {author} {\bibfnamefont {E.~Y.}\ \bibnamefont {Tsymbal}},\ }\href@noop {} {\bibfield  {journal} {\bibinfo  {journal} {Nat. Commun.}\ }\textbf {\bibinfo {volume} {12}},\ \bibinfo {pages} {7061} (\bibinfo {year} {2021}{\natexlab{a}})}\BibitemShut {NoStop}%
\bibitem [{\citenamefont {Šmejkal}\ \emph {et~al.}(2022{\natexlab{c}})\citenamefont {Šmejkal}, \citenamefont {Hellenes}, \citenamefont {González-Hernández}, \citenamefont {Sinova},\ and\ \citenamefont {Jungwirth}}]{817}%
  \BibitemOpen
  \bibfield  {author} {\bibinfo {author} {\bibfnamefont {L.}~\bibnamefont {Šmejkal}}, \bibinfo {author} {\bibfnamefont {A.~B.}\ \bibnamefont {Hellenes}}, \bibinfo {author} {\bibfnamefont {R.}~\bibnamefont {González-Hernández}}, \bibinfo {author} {\bibfnamefont {J.}~\bibnamefont {Sinova}},\ and\ \bibinfo {author} {\bibfnamefont {T.}~\bibnamefont {Jungwirth}},\ }\href@noop {} {\bibfield  {journal} {\bibinfo  {journal} {Phys. Rev. X}\ }\textbf {\bibinfo {volume} {12}},\ \bibinfo {pages} {011028} (\bibinfo {year} {2022}{\natexlab{c}})}\BibitemShut {NoStop}%
\bibitem [{\citenamefont {Chen}\ \emph {et~al.}(2014)\citenamefont {Chen}, \citenamefont {Niu},\ and\ \citenamefont {MacDonald}}]{368}%
  \BibitemOpen
  \bibfield  {author} {\bibinfo {author} {\bibfnamefont {H.}~\bibnamefont {Chen}}, \bibinfo {author} {\bibfnamefont {Q.}~\bibnamefont {Niu}},\ and\ \bibinfo {author} {\bibfnamefont {A.~H.}\ \bibnamefont {MacDonald}},\ }\href@noop {} {\bibfield  {journal} {\bibinfo  {journal} {Phys. Rev. Lett.}\ }\textbf {\bibinfo {volume} {112}},\ \bibinfo {pages} {017205} (\bibinfo {year} {2014})}\BibitemShut {NoStop}%
\bibitem [{\citenamefont {Kübler}\ and\ \citenamefont {Felser}(2014)}]{926}%
  \BibitemOpen
  \bibfield  {author} {\bibinfo {author} {\bibfnamefont {J.}~\bibnamefont {Kübler}}\ and\ \bibinfo {author} {\bibfnamefont {C.}~\bibnamefont {Felser}},\ }\href@noop {} {\bibfield  {journal} {\bibinfo  {journal} {EPL (Europhysics Letters)}\ }\textbf {\bibinfo {volume} {108}},\ \bibinfo {pages} {67001} (\bibinfo {year} {2014})}\BibitemShut {NoStop}%
\bibitem [{\citenamefont {Nagaosa}\ \emph {et~al.}(2010)\citenamefont {Nagaosa}, \citenamefont {Sinova}, \citenamefont {Onoda}, \citenamefont {MacDonald},\ and\ \citenamefont {Ong}}]{212}%
  \BibitemOpen
  \bibfield  {author} {\bibinfo {author} {\bibfnamefont {N.}~\bibnamefont {Nagaosa}}, \bibinfo {author} {\bibfnamefont {J.}~\bibnamefont {Sinova}}, \bibinfo {author} {\bibfnamefont {S.}~\bibnamefont {Onoda}}, \bibinfo {author} {\bibfnamefont {A.~H.}\ \bibnamefont {MacDonald}},\ and\ \bibinfo {author} {\bibfnamefont {N.~P.}\ \bibnamefont {Ong}},\ }\href@noop {} {\bibfield  {journal} {\bibinfo  {journal} {Rev. Mod. Phys.}\ }\textbf {\bibinfo {volume} {82}},\ \bibinfo {pages} {1539} (\bibinfo {year} {2010})}\BibitemShut {NoStop}%
\bibitem [{\citenamefont {Nakatsuji}\ \emph {et~al.}(2015)\citenamefont {Nakatsuji}, \citenamefont {Kiyohara},\ and\ \citenamefont {Higo}}]{599}%
  \BibitemOpen
  \bibfield  {author} {\bibinfo {author} {\bibfnamefont {S.}~\bibnamefont {Nakatsuji}}, \bibinfo {author} {\bibfnamefont {N.}~\bibnamefont {Kiyohara}},\ and\ \bibinfo {author} {\bibfnamefont {T.}~\bibnamefont {Higo}},\ }\href@noop {} {\bibfield  {journal} {\bibinfo  {journal} {Nature}\ }\textbf {\bibinfo {volume} {527}},\ \bibinfo {pages} {212} (\bibinfo {year} {2015})}\BibitemShut {NoStop}%
\bibitem [{\citenamefont {Nayak}\ \emph {et~al.}()\citenamefont {Nayak}, \citenamefont {Fischer}, \citenamefont {Sun}, \citenamefont {Yan}, \citenamefont {Karel}, \citenamefont {Komarek}, \citenamefont {Shekhar}, \citenamefont {Kumar}, \citenamefont {Schnelle}, \citenamefont {Kübler}, \citenamefont {Felser},\ and\ \citenamefont {Parkin}}]{927}%
  \BibitemOpen
  \bibfield  {author} {\bibinfo {author} {\bibfnamefont {A.~K.}\ \bibnamefont {Nayak}}, \bibinfo {author} {\bibfnamefont {J.~E.}\ \bibnamefont {Fischer}}, \bibinfo {author} {\bibfnamefont {Y.}~\bibnamefont {Sun}}, \bibinfo {author} {\bibfnamefont {B.}~\bibnamefont {Yan}}, \bibinfo {author} {\bibfnamefont {J.}~\bibnamefont {Karel}}, \bibinfo {author} {\bibfnamefont {A.~C.}\ \bibnamefont {Komarek}}, \bibinfo {author} {\bibfnamefont {C.}~\bibnamefont {Shekhar}}, \bibinfo {author} {\bibfnamefont {N.}~\bibnamefont {Kumar}}, \bibinfo {author} {\bibfnamefont {W.}~\bibnamefont {Schnelle}}, \bibinfo {author} {\bibfnamefont {J.}~\bibnamefont {Kübler}}, \bibinfo {author} {\bibfnamefont {C.}~\bibnamefont {Felser}},\ and\ \bibinfo {author} {\bibfnamefont {S.~S.~P.}\ \bibnamefont {Parkin}},\ }\href@noop {} {\bibfield  {journal} {\bibinfo  {journal} {Sci. Adv.}\ }\textbf {\bibinfo {volume} {2}},\ \bibinfo {pages} {e1501870}}\BibitemShut {NoStop}%
\bibitem [{\citenamefont {Šmejkal}\ \emph {et~al.}(2022{\natexlab{d}})\citenamefont {Šmejkal}, \citenamefont {MacDonald}, \citenamefont {Sinova}, \citenamefont {Nakatsuji},\ and\ \citenamefont {Jungwirth}}]{480}%
  \BibitemOpen
  \bibfield  {author} {\bibinfo {author} {\bibfnamefont {L.}~\bibnamefont {Šmejkal}}, \bibinfo {author} {\bibfnamefont {A.~H.}\ \bibnamefont {MacDonald}}, \bibinfo {author} {\bibfnamefont {J.}~\bibnamefont {Sinova}}, \bibinfo {author} {\bibfnamefont {S.}~\bibnamefont {Nakatsuji}},\ and\ \bibinfo {author} {\bibfnamefont {T.}~\bibnamefont {Jungwirth}},\ }\href@noop {} {\bibfield  {journal} {\bibinfo  {journal} {Nat. Rev. Mater.}\ }\textbf {\bibinfo {volume} {7}},\ \bibinfo {pages} {482} (\bibinfo {year} {2022}{\natexlab{d}})}\BibitemShut {NoStop}%
\bibitem [{\citenamefont {Suzuki}\ \emph {et~al.}(2017)\citenamefont {Suzuki}, \citenamefont {Koretsune}, \citenamefont {Ochi},\ and\ \citenamefont {Arita}}]{679}%
  \BibitemOpen
  \bibfield  {author} {\bibinfo {author} {\bibfnamefont {M.~T.}\ \bibnamefont {Suzuki}}, \bibinfo {author} {\bibfnamefont {T.}~\bibnamefont {Koretsune}}, \bibinfo {author} {\bibfnamefont {M.}~\bibnamefont {Ochi}},\ and\ \bibinfo {author} {\bibfnamefont {R.}~\bibnamefont {Arita}},\ }\href@noop {} {\bibfield  {journal} {\bibinfo  {journal} {Phys. Rev. B}\ }\textbf {\bibinfo {volume} {95}},\ \bibinfo {pages} {094406} (\bibinfo {year} {2017})}\BibitemShut {NoStop}%
\bibitem [{\citenamefont {Baltz}\ \emph {et~al.}(2018)\citenamefont {Baltz}, \citenamefont {Manchon}, \citenamefont {Tsoi}, \citenamefont {Moriyama}, \citenamefont {Ono},\ and\ \citenamefont {Tserkovnyak}}]{370}%
  \BibitemOpen
  \bibfield  {author} {\bibinfo {author} {\bibfnamefont {V.}~\bibnamefont {Baltz}}, \bibinfo {author} {\bibfnamefont {A.}~\bibnamefont {Manchon}}, \bibinfo {author} {\bibfnamefont {M.}~\bibnamefont {Tsoi}}, \bibinfo {author} {\bibfnamefont {T.}~\bibnamefont {Moriyama}}, \bibinfo {author} {\bibfnamefont {T.}~\bibnamefont {Ono}},\ and\ \bibinfo {author} {\bibfnamefont {Y.}~\bibnamefont {Tserkovnyak}},\ }\href@noop {} {\bibfield  {journal} {\bibinfo  {journal} {Rev. Mod. Phys.}\ }\textbf {\bibinfo {volume} {90}},\ \bibinfo {pages} {015005} (\bibinfo {year} {2018})}\BibitemShut {NoStop}%
\bibitem [{\citenamefont {Jungwirth}\ \emph {et~al.}(2016)\citenamefont {Jungwirth}, \citenamefont {Marti}, \citenamefont {Wadley},\ and\ \citenamefont {Wunderlich}}]{483}%
  \BibitemOpen
  \bibfield  {author} {\bibinfo {author} {\bibfnamefont {T.}~\bibnamefont {Jungwirth}}, \bibinfo {author} {\bibfnamefont {X.}~\bibnamefont {Marti}}, \bibinfo {author} {\bibfnamefont {P.}~\bibnamefont {Wadley}},\ and\ \bibinfo {author} {\bibfnamefont {J.}~\bibnamefont {Wunderlich}},\ }\href@noop {} {\bibfield  {journal} {\bibinfo  {journal} {Nat. Nanotechnol.}\ }\textbf {\bibinfo {volume} {11}},\ \bibinfo {pages} {231} (\bibinfo {year} {2016})}\BibitemShut {NoStop}%
\bibitem [{\citenamefont {Feng}\ \emph {et~al.}(2024)\citenamefont {Feng}, \citenamefont {Wu}, \citenamefont {Wang}, \citenamefont {Gao}, \citenamefont {Ang}, \citenamefont {Zhao}, \citenamefont {Xiao},\ and\ \citenamefont {Yang}}]{1059}%
  \BibitemOpen
  \bibfield  {author} {\bibinfo {author} {\bibfnamefont {X.}~\bibnamefont {Feng}}, \bibinfo {author} {\bibfnamefont {W.}~\bibnamefont {Wu}}, \bibinfo {author} {\bibfnamefont {H.}~\bibnamefont {Wang}}, \bibinfo {author} {\bibfnamefont {W.}~\bibnamefont {Gao}}, \bibinfo {author} {\bibfnamefont {L.~K.}\ \bibnamefont {Ang}}, \bibinfo {author} {\bibfnamefont {Y.}~\bibnamefont {Zhao}}, \bibinfo {author} {\bibfnamefont {C.}~\bibnamefont {Xiao}},\ and\ \bibinfo {author} {\bibfnamefont {S.~A.}\ \bibnamefont {Yang}},\ }\href@noop {} {\bibfield  {journal} {\bibinfo  {journal} {arXiv:2402.00532}\ } (\bibinfo {year} {2024})}\BibitemShut {NoStop}%
\bibitem [{\citenamefont {Xiao}\ \emph {et~al.}(2022)\citenamefont {Xiao}, \citenamefont {Liu}, \citenamefont {Wu}, \citenamefont {Wang}, \citenamefont {Niu},\ and\ \citenamefont {Yang}}]{208}%
  \BibitemOpen
  \bibfield  {author} {\bibinfo {author} {\bibfnamefont {C.}~\bibnamefont {Xiao}}, \bibinfo {author} {\bibfnamefont {H.}~\bibnamefont {Liu}}, \bibinfo {author} {\bibfnamefont {W.}~\bibnamefont {Wu}}, \bibinfo {author} {\bibfnamefont {H.}~\bibnamefont {Wang}}, \bibinfo {author} {\bibfnamefont {Q.}~\bibnamefont {Niu}},\ and\ \bibinfo {author} {\bibfnamefont {S.~A.}\ \bibnamefont {Yang}},\ }\href@noop {} {\bibfield  {journal} {\bibinfo  {journal} {Phys. Rev. Lett.}\ }\textbf {\bibinfo {volume} {129}},\ \bibinfo {pages} {086602} (\bibinfo {year} {2022})}\BibitemShut {NoStop}%
\bibitem [{\citenamefont {Xiao}\ \emph {et~al.}(2021)\citenamefont {Xiao}, \citenamefont {Xiong},\ and\ \citenamefont {Niu}}]{1053}%
  \BibitemOpen
  \bibfield  {author} {\bibinfo {author} {\bibfnamefont {C.}~\bibnamefont {Xiao}}, \bibinfo {author} {\bibfnamefont {B.}~\bibnamefont {Xiong}},\ and\ \bibinfo {author} {\bibfnamefont {Q.}~\bibnamefont {Niu}},\ }\href@noop {} {\bibfield  {journal} {\bibinfo  {journal} {Phys. Rev. B}\ }\textbf {\bibinfo {volume} {104}},\ \bibinfo {pages} {064433} (\bibinfo {year} {2021})}\BibitemShut {NoStop}%
\bibitem [{\citenamefont {Han}\ \emph {et~al.}(2024)\citenamefont {Han}, \citenamefont {Fu}, \citenamefont {Peng}, \citenamefont {Cheng}, \citenamefont {Dai}, \citenamefont {Liu}, \citenamefont {Li}, \citenamefont {Zhang}, \citenamefont {Zhu}, \citenamefont {Bai}, \citenamefont {Zhou}, \citenamefont {Liang}, \citenamefont {Chen}, \citenamefont {Wang}, \citenamefont {Chen}, \citenamefont {Yang}, \citenamefont {Zhang}, \citenamefont {Song}, \citenamefont {Liu},\ and\ \citenamefont {Pan}}]{845}%
  \BibitemOpen
  \bibfield  {author} {\bibinfo {author} {\bibfnamefont {L.}~\bibnamefont {Han}}, \bibinfo {author} {\bibfnamefont {X.}~\bibnamefont {Fu}}, \bibinfo {author} {\bibfnamefont {R.}~\bibnamefont {Peng}}, \bibinfo {author} {\bibfnamefont {X.}~\bibnamefont {Cheng}}, \bibinfo {author} {\bibfnamefont {J.}~\bibnamefont {Dai}}, \bibinfo {author} {\bibfnamefont {L.}~\bibnamefont {Liu}}, \bibinfo {author} {\bibfnamefont {Y.}~\bibnamefont {Li}}, \bibinfo {author} {\bibfnamefont {Y.}~\bibnamefont {Zhang}}, \bibinfo {author} {\bibfnamefont {W.}~\bibnamefont {Zhu}}, \bibinfo {author} {\bibfnamefont {H.}~\bibnamefont {Bai}}, \bibinfo {author} {\bibfnamefont {Y.}~\bibnamefont {Zhou}}, \bibinfo {author} {\bibfnamefont {S.}~\bibnamefont {Liang}}, \bibinfo {author} {\bibfnamefont {C.}~\bibnamefont {Chen}}, \bibinfo {author} {\bibfnamefont {Q.}~\bibnamefont {Wang}}, \bibinfo {author} {\bibfnamefont {X.}~\bibnamefont {Chen}}, \bibinfo {author} {\bibfnamefont {L.}~\bibnamefont {Yang}}, \bibinfo {author} {\bibfnamefont {Y.}~\bibnamefont {Zhang}}, \bibinfo {author} {\bibfnamefont {C.}~\bibnamefont {Song}}, \bibinfo {author} {\bibfnamefont {J.}~\bibnamefont {Liu}},\ and\ \bibinfo {author} {\bibfnamefont {F.}~\bibnamefont {Pan}},\ }\href@noop {} {\bibfield  {journal} {\bibinfo  {journal} {Sci. Adv.}\ }\textbf {\bibinfo {volume} {10}},\ \bibinfo {pages} {eadn0479} (\bibinfo {year} {2024})}\BibitemShut {NoStop}%
\bibitem [{\citenamefont {Scott}(2007)}]{1006}%
  \BibitemOpen
  \bibfield  {author} {\bibinfo {author} {\bibfnamefont {J.~F.}\ \bibnamefont {Scott}},\ }\href@noop {} {\bibfield  {journal} {\bibinfo  {journal} {Nat. Mater.}\ }\textbf {\bibinfo {volume} {6}},\ \bibinfo {pages} {256} (\bibinfo {year} {2007})}\BibitemShut {NoStop}%
\bibitem [{\citenamefont {Tsymbal}(2011)}]{891}%
  \BibitemOpen
  \bibfield  {author} {\bibinfo {author} {\bibfnamefont {E.~Y.}\ \bibnamefont {Tsymbal}},\ }\href@noop {} {\bibfield  {journal} {\bibinfo  {journal} {Nat. Mater.}\ }\textbf {\bibinfo {volume} {11}},\ \bibinfo {pages} {12–13} (\bibinfo {year} {2011})}\BibitemShut {NoStop}%
\bibitem [{\citenamefont {Wadley}\ \emph {et~al.}(2016)\citenamefont {Wadley}, \citenamefont {Howells}, \citenamefont {Železný}, \citenamefont {Andrews}, \citenamefont {Hills}, \citenamefont {Campion}, \citenamefont {Novák}, \citenamefont {Olejník}, \citenamefont {Maccherozzi}, \citenamefont {Dhesi}, \citenamefont {Martin}, \citenamefont {Wagner}, \citenamefont {Wunderlich}, \citenamefont {Freimuth}, \citenamefont {Mokrousov}, \citenamefont {Kuneš}, \citenamefont {Chauhan}, \citenamefont {Grzybowski}, \citenamefont {Rushforth}, \citenamefont {Edmonds}, \citenamefont {Gallagher},\ and\ \citenamefont {Jungwirth}}]{893}%
  \BibitemOpen
  \bibfield  {author} {\bibinfo {author} {\bibfnamefont {P.}~\bibnamefont {Wadley}}, \bibinfo {author} {\bibfnamefont {B.}~\bibnamefont {Howells}}, \bibinfo {author} {\bibfnamefont {J.}~\bibnamefont {Železný}}, \bibinfo {author} {\bibfnamefont {C.}~\bibnamefont {Andrews}}, \bibinfo {author} {\bibfnamefont {V.}~\bibnamefont {Hills}}, \bibinfo {author} {\bibfnamefont {R.~P.}\ \bibnamefont {Campion}}, \bibinfo {author} {\bibfnamefont {V.}~\bibnamefont {Novák}}, \bibinfo {author} {\bibfnamefont {K.}~\bibnamefont {Olejník}}, \bibinfo {author} {\bibfnamefont {F.}~\bibnamefont {Maccherozzi}}, \bibinfo {author} {\bibfnamefont {S.~S.}\ \bibnamefont {Dhesi}}, \bibinfo {author} {\bibfnamefont {S.~Y.}\ \bibnamefont {Martin}}, \bibinfo {author} {\bibfnamefont {T.}~\bibnamefont {Wagner}}, \bibinfo {author} {\bibfnamefont {J.}~\bibnamefont {Wunderlich}}, \bibinfo {author} {\bibfnamefont {F.}~\bibnamefont {Freimuth}}, \bibinfo {author} {\bibfnamefont {Y.}~\bibnamefont {Mokrousov}}, \bibinfo {author} {\bibfnamefont {J.}~\bibnamefont {Kuneš}}, \bibinfo {author} {\bibfnamefont {J.~S.}\ \bibnamefont {Chauhan}}, \bibinfo {author} {\bibfnamefont {M.~J.}\ \bibnamefont {Grzybowski}}, \bibinfo {author} {\bibfnamefont {A.~W.}\ \bibnamefont {Rushforth}}, \bibinfo {author} {\bibfnamefont {K.~W.}\ \bibnamefont {Edmonds}}, \bibinfo {author} {\bibfnamefont {B.~L.}\ \bibnamefont {Gallagher}},\ and\ \bibinfo {author} {\bibfnamefont {T.}~\bibnamefont {Jungwirth}},\ }\href@noop {} {\bibfield  {journal} {\bibinfo  {journal} {Science}\ }\textbf {\bibinfo {volume} {351}},\ \bibinfo {pages} {587} (\bibinfo {year} {2016})}\BibitemShut {NoStop}%
\bibitem [{\citenamefont {Wang}\ \emph {et~al.}(2013)\citenamefont {Wang}, \citenamefont {Alzate},\ and\ \citenamefont {Khalili~Amiri}}]{1005}%
  \BibitemOpen
  \bibfield  {author} {\bibinfo {author} {\bibfnamefont {K.~L.}\ \bibnamefont {Wang}}, \bibinfo {author} {\bibfnamefont {J.~G.}\ \bibnamefont {Alzate}},\ and\ \bibinfo {author} {\bibfnamefont {P.}~\bibnamefont {Khalili~Amiri}},\ }\href@noop {} {\bibfield  {journal} {\bibinfo  {journal} {J. Phys. D: Appl. Phys.}\ }\textbf {\bibinfo {volume} {46}},\ \bibinfo {pages} {074003} (\bibinfo {year} {2013})}\BibitemShut {NoStop}%
\bibitem [{\citenamefont {Tsymbal}(2021)}]{1058}%
  \BibitemOpen
  \bibfield  {author} {\bibinfo {author} {\bibfnamefont {E.~Y.}\ \bibnamefont {Tsymbal}},\ }\href@noop {} {\bibfield  {journal} {\bibinfo  {journal} {Science}\ }\textbf {\bibinfo {volume} {372}},\ \bibinfo {pages} {1389} (\bibinfo {year} {2021})}\BibitemShut {NoStop}%
\bibitem [{\citenamefont {Sun}\ \emph {et~al.}(2024{\natexlab{a}})\citenamefont {Sun}, \citenamefont {Wang}, \citenamefont {Yang}, \citenamefont {Hu}, \citenamefont {Yan}, \citenamefont {Huang},\ and\ \citenamefont {Cheng}}]{730}%
  \BibitemOpen
  \bibfield  {author} {\bibinfo {author} {\bibfnamefont {W.}~\bibnamefont {Sun}}, \bibinfo {author} {\bibfnamefont {W.}~\bibnamefont {Wang}}, \bibinfo {author} {\bibfnamefont {C.}~\bibnamefont {Yang}}, \bibinfo {author} {\bibfnamefont {R.}~\bibnamefont {Hu}}, \bibinfo {author} {\bibfnamefont {S.}~\bibnamefont {Yan}}, \bibinfo {author} {\bibfnamefont {S.}~\bibnamefont {Huang}},\ and\ \bibinfo {author} {\bibfnamefont {Z.}~\bibnamefont {Cheng}},\ }\href@noop {} {\bibfield  {journal} {\bibinfo  {journal} {Nano Lett.}\ }\textbf {\bibinfo {volume} {24}},\ \bibinfo {pages} {11179} (\bibinfo {year} {2024}{\natexlab{a}})}\BibitemShut {NoStop}%
\bibitem [{\citenamefont {He}\ \emph {et~al.}(2023)\citenamefont {He}, \citenamefont {Wang}, \citenamefont {Luo}, \citenamefont {Zeng}, \citenamefont {Chen},\ and\ \citenamefont {Tang}}]{866}%
  \BibitemOpen
  \bibfield  {author} {\bibinfo {author} {\bibfnamefont {R.}~\bibnamefont {He}}, \bibinfo {author} {\bibfnamefont {D.}~\bibnamefont {Wang}}, \bibinfo {author} {\bibfnamefont {N.}~\bibnamefont {Luo}}, \bibinfo {author} {\bibfnamefont {J.}~\bibnamefont {Zeng}}, \bibinfo {author} {\bibfnamefont {K.~Q.}\ \bibnamefont {Chen}},\ and\ \bibinfo {author} {\bibfnamefont {L.~M.}\ \bibnamefont {Tang}},\ }\href@noop {} {\bibfield  {journal} {\bibinfo  {journal} {Phys. Rev. Lett.}\ }\textbf {\bibinfo {volume} {130}},\ \bibinfo {pages} {046401} (\bibinfo {year} {2023})}\BibitemShut {NoStop}%
\bibitem [{\citenamefont {Ji}\ \emph {et~al.}(2023)\citenamefont {Ji}, \citenamefont {Yu}, \citenamefont {Xu},\ and\ \citenamefont {Xiang}}]{128}%
  \BibitemOpen
  \bibfield  {author} {\bibinfo {author} {\bibfnamefont {J.}~\bibnamefont {Ji}}, \bibinfo {author} {\bibfnamefont {G.}~\bibnamefont {Yu}}, \bibinfo {author} {\bibfnamefont {C.}~\bibnamefont {Xu}},\ and\ \bibinfo {author} {\bibfnamefont {H.}~\bibnamefont {Xiang}},\ }\href@noop {} {\bibfield  {journal} {\bibinfo  {journal} {Phys. Rev. Lett.}\ }\textbf {\bibinfo {volume} {130}},\ \bibinfo {pages} {146801} (\bibinfo {year} {2023})}\BibitemShut {NoStop}%
\bibitem [{\citenamefont {Li}\ and\ \citenamefont {Wu}(2017)}]{40}%
  \BibitemOpen
  \bibfield  {author} {\bibinfo {author} {\bibfnamefont {L.}~\bibnamefont {Li}}\ and\ \bibinfo {author} {\bibfnamefont {M.}~\bibnamefont {Wu}},\ }\href@noop {} {\bibfield  {journal} {\bibinfo  {journal} {ACS Nano}\ }\textbf {\bibinfo {volume} {11}},\ \bibinfo {pages} {6382} (\bibinfo {year} {2017})}\BibitemShut {NoStop}%
\bibitem [{\citenamefont {Liu}\ \emph {et~al.}(2024)\citenamefont {Liu}, \citenamefont {Yu},\ and\ \citenamefont {Liu}}]{864}%
  \BibitemOpen
  \bibfield  {author} {\bibinfo {author} {\bibfnamefont {Y.}~\bibnamefont {Liu}}, \bibinfo {author} {\bibfnamefont {J.}~\bibnamefont {Yu}},\ and\ \bibinfo {author} {\bibfnamefont {C.-C.}\ \bibnamefont {Liu}},\ }\href@noop {} {\bibfield  {journal} {\bibinfo  {journal} {Phys. Rev. Lett.}\ }\textbf {\bibinfo {volume} {133}},\ \bibinfo {pages} {206702} (\bibinfo {year} {2024})}\BibitemShut {NoStop}%
\bibitem [{\citenamefont {Pan}\ \emph {et~al.}(2024)\citenamefont {Pan}, \citenamefont {Zhou}, \citenamefont {Lyu}, \citenamefont {Xiao}, \citenamefont {Yang},\ and\ \citenamefont {Sun}}]{882}%
  \BibitemOpen
  \bibfield  {author} {\bibinfo {author} {\bibfnamefont {B.}~\bibnamefont {Pan}}, \bibinfo {author} {\bibfnamefont {P.}~\bibnamefont {Zhou}}, \bibinfo {author} {\bibfnamefont {P.}~\bibnamefont {Lyu}}, \bibinfo {author} {\bibfnamefont {H.}~\bibnamefont {Xiao}}, \bibinfo {author} {\bibfnamefont {X.}~\bibnamefont {Yang}},\ and\ \bibinfo {author} {\bibfnamefont {L.}~\bibnamefont {Sun}},\ }\href@noop {} {\bibfield  {journal} {\bibinfo  {journal} {Phys. Rev. Lett.}\ }\textbf {\bibinfo {volume} {133}},\ \bibinfo {pages} {166701} (\bibinfo {year} {2024})}\BibitemShut {NoStop}%
\bibitem [{\citenamefont {Zeng}\ and\ \citenamefont {Zhao}(2024{\natexlab{a}})}]{861}%
  \BibitemOpen
  \bibfield  {author} {\bibinfo {author} {\bibfnamefont {S.}~\bibnamefont {Zeng}}\ and\ \bibinfo {author} {\bibfnamefont {Y.-J.}\ \bibnamefont {Zhao}},\ }\href@noop {} {\bibfield  {journal} {\bibinfo  {journal} {Phys. Rev. B}\ }\textbf {\bibinfo {volume} {110}},\ \bibinfo {pages} {174410} (\bibinfo {year} {2024}{\natexlab{a}})}\BibitemShut {NoStop}%
\bibitem [{\citenamefont {Sun}\ \emph {et~al.}(2024{\natexlab{b}})\citenamefont {Sun}, \citenamefont {Ye}, \citenamefont {Liang}, \citenamefont {Ding}, \citenamefont {Dong},\ and\ \citenamefont {Wang}}]{1088}%
  \BibitemOpen
  \bibfield  {author} {\bibinfo {author} {\bibfnamefont {W.}~\bibnamefont {Sun}}, \bibinfo {author} {\bibfnamefont {H.}~\bibnamefont {Ye}}, \bibinfo {author} {\bibfnamefont {L.}~\bibnamefont {Liang}}, \bibinfo {author} {\bibfnamefont {N.}~\bibnamefont {Ding}}, \bibinfo {author} {\bibfnamefont {S.}~\bibnamefont {Dong}},\ and\ \bibinfo {author} {\bibfnamefont {S.-S.}\ \bibnamefont {Wang}},\ }\href@noop {} {\bibfield  {journal} {\bibinfo  {journal} {Phys. Rev. B}\ }\textbf {\bibinfo {volume} {110}},\ \bibinfo {pages} {224418} (\bibinfo {year} {2024}{\natexlab{b}})}\BibitemShut {NoStop}%
\bibitem [{\citenamefont {Bradley}\ and\ \citenamefont {Cracknell}(2010)}]{304}%
  \BibitemOpen
  \bibfield  {author} {\bibinfo {author} {\bibfnamefont {C.}~\bibnamefont {Bradley}}\ and\ \bibinfo {author} {\bibfnamefont {A.}~\bibnamefont {Cracknell}},\ }\href@noop {} {\emph {\bibinfo {title} {The mathematical theory of symmetry in solids: representation theory for point groups and space groups}}}\ (\bibinfo  {publisher} {Oxford University Press},\ \bibinfo {year} {2010})\BibitemShut {NoStop}%
\bibitem [{\citenamefont {Brinkman}\ and\ \citenamefont {Elliott}(1966{\natexlab{a}})}]{1009}%
  \BibitemOpen
  \bibfield  {author} {\bibinfo {author} {\bibfnamefont {W.}~\bibnamefont {Brinkman}}\ and\ \bibinfo {author} {\bibfnamefont {R.~J.}\ \bibnamefont {Elliott}},\ }\href@noop {} {\bibfield  {journal} {\bibinfo  {journal} {Proceedings of the Royal Society of London. Series A. Mathematical and Physical Sciences}\ }\textbf {\bibinfo {volume} {294}},\ \bibinfo {pages} {343} (\bibinfo {year} {1966}{\natexlab{a}})}\BibitemShut {NoStop}%
\bibitem [{\citenamefont {Brinkman}\ and\ \citenamefont {Elliott}(1966{\natexlab{b}})}]{1007}%
  \BibitemOpen
  \bibfield  {author} {\bibinfo {author} {\bibfnamefont {W.}~\bibnamefont {Brinkman}}\ and\ \bibinfo {author} {\bibfnamefont {R.~J.}\ \bibnamefont {Elliott}},\ }\href@noop {} {\bibfield  {journal} {\bibinfo  {journal} {J. Appl. Phys.}\ }\textbf {\bibinfo {volume} {37}},\ \bibinfo {pages} {1457} (\bibinfo {year} {1966}{\natexlab{b}})}\BibitemShut {NoStop}%
\bibitem [{\citenamefont {Litvin}\ and\ \citenamefont {Opechowski}(1974)}]{1011}%
  \BibitemOpen
  \bibfield  {author} {\bibinfo {author} {\bibfnamefont {D.~B.}\ \bibnamefont {Litvin}}\ and\ \bibinfo {author} {\bibfnamefont {W.}~\bibnamefont {Opechowski}},\ }\href@noop {} {\bibfield  {journal} {\bibinfo  {journal} {Physica}\ }\textbf {\bibinfo {volume} {76}},\ \bibinfo {pages} {538} (\bibinfo {year} {1974})}\BibitemShut {NoStop}%
\bibitem [{\citenamefont {Zhu}\ \emph {et~al.}(2025)\citenamefont {Zhu}, \citenamefont {Sun}, \citenamefont {Pan}, \citenamefont {Deng},\ and\ \citenamefont {Du}}]{995}%
  \BibitemOpen
  \bibfield  {author} {\bibinfo {author} {\bibfnamefont {Y.}~\bibnamefont {Zhu}}, \bibinfo {author} {\bibfnamefont {J.-T.}\ \bibnamefont {Sun}}, \bibinfo {author} {\bibfnamefont {J.}~\bibnamefont {Pan}}, \bibinfo {author} {\bibfnamefont {J.}~\bibnamefont {Deng}},\ and\ \bibinfo {author} {\bibfnamefont {S.}~\bibnamefont {Du}},\ }\href@noop {} {\bibfield  {journal} {\bibinfo  {journal} {Phys. Rev. Lett.}\ }\textbf {\bibinfo {volume} {134}},\ \bibinfo {pages} {046403} (\bibinfo {year} {2025})}\BibitemShut {NoStop}%
\bibitem [{\citenamefont {Shao}\ \emph {et~al.}(2021{\natexlab{b}})\citenamefont {Shao}, \citenamefont {Ding}, \citenamefont {Gurung}, \citenamefont {Zhang},\ and\ \citenamefont {Tsymbal}}]{849}%
  \BibitemOpen
  \bibfield  {author} {\bibinfo {author} {\bibfnamefont {D.-F.}\ \bibnamefont {Shao}}, \bibinfo {author} {\bibfnamefont {J.}~\bibnamefont {Ding}}, \bibinfo {author} {\bibfnamefont {G.}~\bibnamefont {Gurung}}, \bibinfo {author} {\bibfnamefont {S.-H.}\ \bibnamefont {Zhang}},\ and\ \bibinfo {author} {\bibfnamefont {E.~Y.}\ \bibnamefont {Tsymbal}},\ }\href@noop {} {\bibfield  {journal} {\bibinfo  {journal} {Phys. Rev. Appl.}\ }\textbf {\bibinfo {volume} {15}},\ \bibinfo {pages} {024057} (\bibinfo {year} {2021}{\natexlab{b}})}\BibitemShut {NoStop}%
\bibitem [{\citenamefont {Cao}\ \emph {et~al.}(2023)\citenamefont {Cao}, \citenamefont {Shao}, \citenamefont {Huang}, \citenamefont {Gurung},\ and\ \citenamefont {Tsymbal}}]{343}%
  \BibitemOpen
  \bibfield  {author} {\bibinfo {author} {\bibfnamefont {T.}~\bibnamefont {Cao}}, \bibinfo {author} {\bibfnamefont {D.-F.}\ \bibnamefont {Shao}}, \bibinfo {author} {\bibfnamefont {K.}~\bibnamefont {Huang}}, \bibinfo {author} {\bibfnamefont {G.}~\bibnamefont {Gurung}},\ and\ \bibinfo {author} {\bibfnamefont {E.~Y.}\ \bibnamefont {Tsymbal}},\ }\href@noop {} {\bibfield  {journal} {\bibinfo  {journal} {Nano Lett.}\ }\textbf {\bibinfo {volume} {23}},\ \bibinfo {pages} {3781} (\bibinfo {year} {2023})}\BibitemShut {NoStop}%
\bibitem [{101()}]{1013}%
  \BibitemOpen
  \href@noop {} {\bibinfo  {journal} {See Supplemental Material for details about group theory analysis, computational methods and the DFT results of bilayer AgF\textsubscript{2} and CrI\textsubscript{3}, which includes Refs. \cite{408, 407, 789, 413, 139, 591, 618, 401, 904, 411, 128, 792, 395, 394, 1060, 410, 205, 406, 393, 746, 905, 796, 751, 1086, 409, 268, 995}}\ }\BibitemShut {NoStop}%
\bibitem [{\citenamefont {Sødequist}\ and\ \citenamefont {Olsen}(2024)}]{751}%
  \BibitemOpen
\bibfield  {journal} {  }\bibfield  {author} {\bibinfo {author} {\bibfnamefont {J.}~\bibnamefont {Sødequist}}\ and\ \bibinfo {author} {\bibfnamefont {T.}~\bibnamefont {Olsen}},\ }\href@noop {} {\bibfield  {journal} {\bibinfo  {journal} {Appl. Phys. Lett.}\ }\textbf {\bibinfo {volume} {124}},\ \bibinfo {pages} {182409} (\bibinfo {year} {2024})}\BibitemShut {NoStop}%
\bibitem [{\citenamefont {Zeng}\ and\ \citenamefont {Zhao}(2024{\natexlab{b}})}]{756}%
  \BibitemOpen
  \bibfield  {author} {\bibinfo {author} {\bibfnamefont {S.}~\bibnamefont {Zeng}}\ and\ \bibinfo {author} {\bibfnamefont {Y.-J.}\ \bibnamefont {Zhao}},\ }\href@noop {} {\bibfield  {journal} {\bibinfo  {journal} {Phys. Rev. B}\ }\textbf {\bibinfo {volume} {110}},\ \bibinfo {pages} {054406} (\bibinfo {year} {2024}{\natexlab{b}})}\BibitemShut {NoStop}%
\bibitem [{\citenamefont {Liu}\ \emph {et~al.}(2023)\citenamefont {Liu}, \citenamefont {Li},\ and\ \citenamefont {Liu}}]{465}%
  \BibitemOpen
  \bibfield  {author} {\bibinfo {author} {\bibfnamefont {Y.}~\bibnamefont {Liu}}, \bibinfo {author} {\bibfnamefont {J.}~\bibnamefont {Li}},\ and\ \bibinfo {author} {\bibfnamefont {Q.}~\bibnamefont {Liu}},\ }\href@noop {} {\bibfield  {journal} {\bibinfo  {journal} {Nano Lett.}\ }\textbf {\bibinfo {volume} {23}},\ \bibinfo {pages} {8650} (\bibinfo {year} {2023})}\BibitemShut {NoStop}%
\bibitem [{\citenamefont {Li}\ \emph {et~al.}(2025)\citenamefont {Li}, \citenamefont {Zhang}, \citenamefont {Lu}, \citenamefont {Shao}, \citenamefont {Bao}, \citenamefont {Zheng}, \citenamefont {Tong},\ and\ \citenamefont {Duan}}]{1060}%
  \BibitemOpen
  \bibfield  {author} {\bibinfo {author} {\bibfnamefont {Y.-Q.}\ \bibnamefont {Li}}, \bibinfo {author} {\bibfnamefont {Y.-K.}\ \bibnamefont {Zhang}}, \bibinfo {author} {\bibfnamefont {X.-L.}\ \bibnamefont {Lu}}, \bibinfo {author} {\bibfnamefont {Y.-P.}\ \bibnamefont {Shao}}, \bibinfo {author} {\bibfnamefont {Z.-Q.}\ \bibnamefont {Bao}}, \bibinfo {author} {\bibfnamefont {J.-D.}\ \bibnamefont {Zheng}}, \bibinfo {author} {\bibfnamefont {W.-Y.}\ \bibnamefont {Tong}},\ and\ \bibinfo {author} {\bibfnamefont {C.-G.}\ \bibnamefont {Duan}},\ }\href@noop {} {\bibfield  {journal} {\bibinfo  {journal} {Nano Lett.}\ }\textbf {\bibinfo {volume} {25}},\ \bibinfo {pages} {6032} (\bibinfo {year} {2025})}\BibitemShut {NoStop}%
\bibitem [{\citenamefont {Ni}\ \emph {et~al.}(2021)\citenamefont {Ni}, \citenamefont {Haglund}, \citenamefont {Wang}, \citenamefont {Xu}, \citenamefont {Bernhard}, \citenamefont {Mandrus}, \citenamefont {Qian}, \citenamefont {Mele}, \citenamefont {Kane},\ and\ \citenamefont {Wu}}]{205}%
  \BibitemOpen
  \bibfield  {author} {\bibinfo {author} {\bibfnamefont {Z.}~\bibnamefont {Ni}}, \bibinfo {author} {\bibfnamefont {A.~V.}\ \bibnamefont {Haglund}}, \bibinfo {author} {\bibfnamefont {H.}~\bibnamefont {Wang}}, \bibinfo {author} {\bibfnamefont {B.}~\bibnamefont {Xu}}, \bibinfo {author} {\bibfnamefont {C.}~\bibnamefont {Bernhard}}, \bibinfo {author} {\bibfnamefont {D.~G.}\ \bibnamefont {Mandrus}}, \bibinfo {author} {\bibfnamefont {X.}~\bibnamefont {Qian}}, \bibinfo {author} {\bibfnamefont {E.~J.}\ \bibnamefont {Mele}}, \bibinfo {author} {\bibfnamefont {C.~L.}\ \bibnamefont {Kane}},\ and\ \bibinfo {author} {\bibfnamefont {L.}~\bibnamefont {Wu}},\ }\href@noop {} {\bibfield  {journal} {\bibinfo  {journal} {Nat. Nanotechnol.}\ }\textbf {\bibinfo {volume} {16}},\ \bibinfo {pages} {782} (\bibinfo {year} {2021})}\BibitemShut {NoStop}%
\bibitem [{\citenamefont {Fischer}\ \emph {et~al.}(1971)\citenamefont {Fischer}, \citenamefont {Schwarzenbach},\ and\ \citenamefont {Rietveld}}]{901}%
  \BibitemOpen
  \bibfield  {author} {\bibinfo {author} {\bibfnamefont {P.}~\bibnamefont {Fischer}}, \bibinfo {author} {\bibfnamefont {D.}~\bibnamefont {Schwarzenbach}},\ and\ \bibinfo {author} {\bibfnamefont {H.~M.}\ \bibnamefont {Rietveld}},\ }\href@noop {} {\bibfield  {journal} {\bibinfo  {journal} {J. Phys. Chem. Solids}\ }\textbf {\bibinfo {volume} {32}},\ \bibinfo {pages} {543} (\bibinfo {year} {1971})}\BibitemShut {NoStop}%
\bibitem [{\citenamefont {Xu}\ \emph {et~al.}(2020)\citenamefont {Xu}, \citenamefont {Ma}, \citenamefont {Zhang}, \citenamefont {Lei}, \citenamefont {Huang},\ and\ \citenamefont {Dai}}]{781}%
  \BibitemOpen
  \bibfield  {author} {\bibinfo {author} {\bibfnamefont {X.}~\bibnamefont {Xu}}, \bibinfo {author} {\bibfnamefont {Y.}~\bibnamefont {Ma}}, \bibinfo {author} {\bibfnamefont {T.}~\bibnamefont {Zhang}}, \bibinfo {author} {\bibfnamefont {C.}~\bibnamefont {Lei}}, \bibinfo {author} {\bibfnamefont {B.}~\bibnamefont {Huang}},\ and\ \bibinfo {author} {\bibfnamefont {Y.}~\bibnamefont {Dai}},\ }\href@noop {} {\bibfield  {journal} {\bibinfo  {journal} {Nanoscale Horiz.}\ }\textbf {\bibinfo {volume} {5}},\ \bibinfo {pages} {1386} (\bibinfo {year} {2020})}\BibitemShut {NoStop}%
\bibitem [{\citenamefont {Finley}\ and\ \citenamefont {Liu}(2020)}]{850}%
  \BibitemOpen
  \bibfield  {author} {\bibinfo {author} {\bibfnamefont {J.}~\bibnamefont {Finley}}\ and\ \bibinfo {author} {\bibfnamefont {L.}~\bibnamefont {Liu}},\ }\href@noop {} {\bibfield  {journal} {\bibinfo  {journal} {Appl. Phys. Lett.}\ }\textbf {\bibinfo {volume} {116}},\ \bibinfo {pages} {110501} (\bibinfo {year} {2020})}\BibitemShut {NoStop}%
\bibitem [{\citenamefont {Kim}\ \emph {et~al.}(2021)\citenamefont {Kim}, \citenamefont {Beach}, \citenamefont {Lee}, \citenamefont {Ono}, \citenamefont {Rasing},\ and\ \citenamefont {Yang}}]{495}%
  \BibitemOpen
  \bibfield  {author} {\bibinfo {author} {\bibfnamefont {S.~K.}\ \bibnamefont {Kim}}, \bibinfo {author} {\bibfnamefont {G.~S.~D.}\ \bibnamefont {Beach}}, \bibinfo {author} {\bibfnamefont {K.-J.}\ \bibnamefont {Lee}}, \bibinfo {author} {\bibfnamefont {T.}~\bibnamefont {Ono}}, \bibinfo {author} {\bibfnamefont {T.}~\bibnamefont {Rasing}},\ and\ \bibinfo {author} {\bibfnamefont {H.}~\bibnamefont {Yang}},\ }\href@noop {} {\bibfield  {journal} {\bibinfo  {journal} {Nat. Mater.}\ }\textbf {\bibinfo {volume} {21}},\ \bibinfo {pages} {24} (\bibinfo {year} {2021})}\BibitemShut {NoStop}%
\bibitem [{\citenamefont {Liu}\ \emph {et~al.}(2025{\natexlab{b}})\citenamefont {Liu}, \citenamefont {Guo}, \citenamefont {Li},\ and\ \citenamefont {Liu}}]{1052}%
  \BibitemOpen
  \bibfield  {author} {\bibinfo {author} {\bibfnamefont {Y.}~\bibnamefont {Liu}}, \bibinfo {author} {\bibfnamefont {S.~D.}\ \bibnamefont {Guo}}, \bibinfo {author} {\bibfnamefont {Y.}~\bibnamefont {Li}},\ and\ \bibinfo {author} {\bibfnamefont {C.~C.}\ \bibnamefont {Liu}},\ }\href@noop {} {\bibfield  {journal} {\bibinfo  {journal} {Phys. Rev. Lett.}\ }\textbf {\bibinfo {volume} {134}},\ \bibinfo {pages} {116703} (\bibinfo {year} {2025}{\natexlab{b}})}\BibitemShut {NoStop}%
\bibitem [{\citenamefont {Yuan}\ \emph {et~al.}(2024)\citenamefont {Yuan}, \citenamefont {Georgescu},\ and\ \citenamefont {Rondinelli}}]{873}%
  \BibitemOpen
  \bibfield  {author} {\bibinfo {author} {\bibfnamefont {L.-D.}\ \bibnamefont {Yuan}}, \bibinfo {author} {\bibfnamefont {A.~B.}\ \bibnamefont {Georgescu}},\ and\ \bibinfo {author} {\bibfnamefont {J.~M.}\ \bibnamefont {Rondinelli}},\ }\href@noop {} {\bibfield  {journal} {\bibinfo  {journal} {Phys. Rev. Lett.}\ }\textbf {\bibinfo {volume} {133}},\ \bibinfo {pages} {216701} (\bibinfo {year} {2024})}\BibitemShut {NoStop}%
\bibitem [{\citenamefont {Huang}\ \emph {et~al.}(2017)\citenamefont {Huang}, \citenamefont {Clark}, \citenamefont {Navarro-Moratalla}, \citenamefont {Klein}, \citenamefont {Cheng}, \citenamefont {Seyler}, \citenamefont {Zhong}, \citenamefont {Schmidgall}, \citenamefont {McGuire}, \citenamefont {Cobden}, \citenamefont {Yao}, \citenamefont {Xiao}, \citenamefont {Jarillo-Herrero},\ and\ \citenamefont {Xu}}]{76}%
  \BibitemOpen
  \bibfield  {author} {\bibinfo {author} {\bibfnamefont {B.}~\bibnamefont {Huang}}, \bibinfo {author} {\bibfnamefont {G.}~\bibnamefont {Clark}}, \bibinfo {author} {\bibfnamefont {E.}~\bibnamefont {Navarro-Moratalla}}, \bibinfo {author} {\bibfnamefont {D.~R.}\ \bibnamefont {Klein}}, \bibinfo {author} {\bibfnamefont {R.}~\bibnamefont {Cheng}}, \bibinfo {author} {\bibfnamefont {K.~L.}\ \bibnamefont {Seyler}}, \bibinfo {author} {\bibfnamefont {D.}~\bibnamefont {Zhong}}, \bibinfo {author} {\bibfnamefont {E.}~\bibnamefont {Schmidgall}}, \bibinfo {author} {\bibfnamefont {M.~A.}\ \bibnamefont {McGuire}}, \bibinfo {author} {\bibfnamefont {D.~H.}\ \bibnamefont {Cobden}}, \bibinfo {author} {\bibfnamefont {W.}~\bibnamefont {Yao}}, \bibinfo {author} {\bibfnamefont {D.}~\bibnamefont {Xiao}}, \bibinfo {author} {\bibfnamefont {P.}~\bibnamefont {Jarillo-Herrero}},\ and\ \bibinfo {author} {\bibfnamefont {X.}~\bibnamefont {Xu}},\ }\href@noop {} {\bibfield  {journal} {\bibinfo  {journal} {Nature}\ }\textbf {\bibinfo {volume} {546}},\ \bibinfo {pages} {270} (\bibinfo {year} {2017})}\BibitemShut {NoStop}%
\bibitem [{\citenamefont {Gong}\ \emph {et~al.}(2017)\citenamefont {Gong}, \citenamefont {Li}, \citenamefont {Li}, \citenamefont {Ji}, \citenamefont {Stern}, \citenamefont {Xia}, \citenamefont {Cao}, \citenamefont {Bao}, \citenamefont {Wang}, \citenamefont {Wang}, \citenamefont {Qiu}, \citenamefont {Cava}, \citenamefont {Louie}, \citenamefont {Xia},\ and\ \citenamefont {Zhang}}]{445}%
  \BibitemOpen
  \bibfield  {author} {\bibinfo {author} {\bibfnamefont {C.}~\bibnamefont {Gong}}, \bibinfo {author} {\bibfnamefont {L.}~\bibnamefont {Li}}, \bibinfo {author} {\bibfnamefont {Z.}~\bibnamefont {Li}}, \bibinfo {author} {\bibfnamefont {H.}~\bibnamefont {Ji}}, \bibinfo {author} {\bibfnamefont {A.}~\bibnamefont {Stern}}, \bibinfo {author} {\bibfnamefont {Y.}~\bibnamefont {Xia}}, \bibinfo {author} {\bibfnamefont {T.}~\bibnamefont {Cao}}, \bibinfo {author} {\bibfnamefont {W.}~\bibnamefont {Bao}}, \bibinfo {author} {\bibfnamefont {C.}~\bibnamefont {Wang}}, \bibinfo {author} {\bibfnamefont {Y.}~\bibnamefont {Wang}}, \bibinfo {author} {\bibfnamefont {Z.~Q.}\ \bibnamefont {Qiu}}, \bibinfo {author} {\bibfnamefont {R.~J.}\ \bibnamefont {Cava}}, \bibinfo {author} {\bibfnamefont {S.~G.}\ \bibnamefont {Louie}}, \bibinfo {author} {\bibfnamefont {J.}~\bibnamefont {Xia}},\ and\ \bibinfo {author} {\bibfnamefont {X.}~\bibnamefont {Zhang}},\ }\href@noop {} {\bibfield  {journal} {\bibinfo  {journal} {Nature}\ }\textbf {\bibinfo {volume} {546}},\ \bibinfo {pages} {265} (\bibinfo {year} {2017})}\BibitemShut {NoStop}%
\bibitem [{\citenamefont {Deng}\ \emph {et~al.}(2020)\citenamefont {Deng}, \citenamefont {Yu}, \citenamefont {Shi}, \citenamefont {Guo}, \citenamefont {Xu}, \citenamefont {Wang}, \citenamefont {Chen},\ and\ \citenamefont {Zhang}}]{1069}%
  \BibitemOpen
  \bibfield  {author} {\bibinfo {author} {\bibfnamefont {Y.}~\bibnamefont {Deng}}, \bibinfo {author} {\bibfnamefont {Y.}~\bibnamefont {Yu}}, \bibinfo {author} {\bibfnamefont {M.~Z.}\ \bibnamefont {Shi}}, \bibinfo {author} {\bibfnamefont {Z.}~\bibnamefont {Guo}}, \bibinfo {author} {\bibfnamefont {Z.}~\bibnamefont {Xu}}, \bibinfo {author} {\bibfnamefont {J.}~\bibnamefont {Wang}}, \bibinfo {author} {\bibfnamefont {X.~H.}\ \bibnamefont {Chen}},\ and\ \bibinfo {author} {\bibfnamefont {Y.}~\bibnamefont {Zhang}},\ }\href@noop {} {\bibfield  {journal} {\bibinfo  {journal} {Science}\ }\textbf {\bibinfo {volume} {367}},\ \bibinfo {pages} {895} (\bibinfo {year} {2020})}\BibitemShut {NoStop}%
\bibitem [{\citenamefont {Otrokov}\ \emph {et~al.}(2019)\citenamefont {Otrokov}, \citenamefont {Rusinov}, \citenamefont {Blanco-Rey}, \citenamefont {Hoffmann}, \citenamefont {Vyazovskaya}, \citenamefont {Eremeev}, \citenamefont {Ernst}, \citenamefont {Echenique}, \citenamefont {Arnau},\ and\ \citenamefont {Chulkov}}]{1067}%
  \BibitemOpen
  \bibfield  {author} {\bibinfo {author} {\bibfnamefont {M.~M.}\ \bibnamefont {Otrokov}}, \bibinfo {author} {\bibfnamefont {I.~P.}\ \bibnamefont {Rusinov}}, \bibinfo {author} {\bibfnamefont {M.}~\bibnamefont {Blanco-Rey}}, \bibinfo {author} {\bibfnamefont {M.}~\bibnamefont {Hoffmann}}, \bibinfo {author} {\bibfnamefont {A.~Y.}\ \bibnamefont {Vyazovskaya}}, \bibinfo {author} {\bibfnamefont {S.~V.}\ \bibnamefont {Eremeev}}, \bibinfo {author} {\bibfnamefont {A.}~\bibnamefont {Ernst}}, \bibinfo {author} {\bibfnamefont {P.~M.}\ \bibnamefont {Echenique}}, \bibinfo {author} {\bibfnamefont {A.}~\bibnamefont {Arnau}},\ and\ \bibinfo {author} {\bibfnamefont {E.~V.}\ \bibnamefont {Chulkov}},\ }\href@noop {} {\bibfield  {journal} {\bibinfo  {journal} {Phys. Rev. Lett.}\ }\textbf {\bibinfo {volume} {122}},\ \bibinfo {pages} {107202} (\bibinfo {year} {2019})}\BibitemShut {NoStop}%
\bibitem [{\citenamefont {Zhang}\ \emph {et~al.}(2024)\citenamefont {Zhang}, \citenamefont {Cui}, \citenamefont {Li}, \citenamefont {Duan}, \citenamefont {Li}, \citenamefont {Yu},\ and\ \citenamefont {Yao}}]{706}%
  \BibitemOpen
  \bibfield  {author} {\bibinfo {author} {\bibfnamefont {R.~W.}\ \bibnamefont {Zhang}}, \bibinfo {author} {\bibfnamefont {C.}~\bibnamefont {Cui}}, \bibinfo {author} {\bibfnamefont {R.}~\bibnamefont {Li}}, \bibinfo {author} {\bibfnamefont {J.}~\bibnamefont {Duan}}, \bibinfo {author} {\bibfnamefont {L.}~\bibnamefont {Li}}, \bibinfo {author} {\bibfnamefont {Z.~M.}\ \bibnamefont {Yu}},\ and\ \bibinfo {author} {\bibfnamefont {Y.}~\bibnamefont {Yao}},\ }\href@noop {} {\bibfield  {journal} {\bibinfo  {journal} {Phys. Rev. Lett.}\ }\textbf {\bibinfo {volume} {133}},\ \bibinfo {pages} {056401} (\bibinfo {year} {2024})}\BibitemShut {NoStop}%
\bibitem [{\citenamefont {Duan}\ \emph {et~al.}(2025)\citenamefont {Duan}, \citenamefont {Zhang}, \citenamefont {Zhu}, \citenamefont {Liu}, \citenamefont {Zhang}, \citenamefont {Žutić},\ and\ \citenamefont {Zhou}}]{1055}%
  \BibitemOpen
  \bibfield  {author} {\bibinfo {author} {\bibfnamefont {X.}~\bibnamefont {Duan}}, \bibinfo {author} {\bibfnamefont {J.}~\bibnamefont {Zhang}}, \bibinfo {author} {\bibfnamefont {Z.}~\bibnamefont {Zhu}}, \bibinfo {author} {\bibfnamefont {Y.}~\bibnamefont {Liu}}, \bibinfo {author} {\bibfnamefont {Z.}~\bibnamefont {Zhang}}, \bibinfo {author} {\bibfnamefont {I.}~\bibnamefont {Žutić}},\ and\ \bibinfo {author} {\bibfnamefont {T.}~\bibnamefont {Zhou}},\ }\href@noop {} {\bibfield  {journal} {\bibinfo  {journal} {Phys. Rev. Lett.}\ }\textbf {\bibinfo {volume} {134}},\ \bibinfo {pages} {106801} (\bibinfo {year} {2025})}\BibitemShut {NoStop}%
\bibitem [{\citenamefont {Gu}\ \emph {et~al.}(2025)\citenamefont {Gu}, \citenamefont {Liu}, \citenamefont {Zhu}, \citenamefont {Yananose}, \citenamefont {Chen}, \citenamefont {Hu}, \citenamefont {Stroppa},\ and\ \citenamefont {Liu}}]{1056}%
  \BibitemOpen
  \bibfield  {author} {\bibinfo {author} {\bibfnamefont {M.}~\bibnamefont {Gu}}, \bibinfo {author} {\bibfnamefont {Y.}~\bibnamefont {Liu}}, \bibinfo {author} {\bibfnamefont {H.}~\bibnamefont {Zhu}}, \bibinfo {author} {\bibfnamefont {K.}~\bibnamefont {Yananose}}, \bibinfo {author} {\bibfnamefont {X.}~\bibnamefont {Chen}}, \bibinfo {author} {\bibfnamefont {Y.}~\bibnamefont {Hu}}, \bibinfo {author} {\bibfnamefont {A.}~\bibnamefont {Stroppa}},\ and\ \bibinfo {author} {\bibfnamefont {Q.}~\bibnamefont {Liu}},\ }\href@noop {} {\bibfield  {journal} {\bibinfo  {journal} {Phys. Rev. Lett.}\ }\textbf {\bibinfo {volume} {134}},\ \bibinfo {pages} {106802} (\bibinfo {year} {2025})}\BibitemShut {NoStop}%
\bibitem [{\citenamefont {Aroyo}\ \emph {et~al.}(2006{\natexlab{a}})\citenamefont {Aroyo}, \citenamefont {Kirov}, \citenamefont {Capillas}, \citenamefont {Perez-Mato},\ and\ \citenamefont {Wondratschek}}]{408}%
  \BibitemOpen
  \bibfield  {author} {\bibinfo {author} {\bibfnamefont {M.~I.}\ \bibnamefont {Aroyo}}, \bibinfo {author} {\bibfnamefont {A.}~\bibnamefont {Kirov}}, \bibinfo {author} {\bibfnamefont {C.}~\bibnamefont {Capillas}}, \bibinfo {author} {\bibfnamefont {J.~M.}\ \bibnamefont {Perez-Mato}},\ and\ \bibinfo {author} {\bibfnamefont {H.}~\bibnamefont {Wondratschek}},\ }\href@noop {} {\bibfield  {journal} {\bibinfo  {journal} {Acta Crystallogr A}\ }\textbf {\bibinfo {volume} {62}},\ \bibinfo {pages} {115} (\bibinfo {year} {2006}{\natexlab{a}})}\BibitemShut {NoStop}%
\bibitem [{\citenamefont {Aroyo}\ \emph {et~al.}(2006{\natexlab{b}})\citenamefont {Aroyo}, \citenamefont {Perez-Mato}, \citenamefont {Capillas}, \citenamefont {Kroumova}, \citenamefont {Ivantchev}, \citenamefont {Madariaga}, \citenamefont {Kirov},\ and\ \citenamefont {Wondratschek}}]{407}%
  \BibitemOpen
  \bibfield  {author} {\bibinfo {author} {\bibfnamefont {M.~I.}\ \bibnamefont {Aroyo}}, \bibinfo {author} {\bibfnamefont {J.~M.}\ \bibnamefont {Perez-Mato}}, \bibinfo {author} {\bibfnamefont {C.}~\bibnamefont {Capillas}}, \bibinfo {author} {\bibfnamefont {E.}~\bibnamefont {Kroumova}}, \bibinfo {author} {\bibfnamefont {S.}~\bibnamefont {Ivantchev}}, \bibinfo {author} {\bibfnamefont {G.}~\bibnamefont {Madariaga}}, \bibinfo {author} {\bibfnamefont {A.}~\bibnamefont {Kirov}},\ and\ \bibinfo {author} {\bibfnamefont {H.}~\bibnamefont {Wondratschek}},\ }\href@noop {} {\bibfield  {journal} {\bibinfo  {journal} {Z. Kristallogr. - Cryst. Mater.}\ }\textbf {\bibinfo {volume} {221}},\ \bibinfo {pages} {15} (\bibinfo {year} {2006}{\natexlab{b}})}\BibitemShut {NoStop}%
\bibitem [{\citenamefont {Derzsi}\ \emph {et~al.}(2022)\citenamefont {Derzsi}, \citenamefont {Tokár}, \citenamefont {Piekarz},\ and\ \citenamefont {Grochala}}]{789}%
  \BibitemOpen
  \bibfield  {author} {\bibinfo {author} {\bibfnamefont {M.}~\bibnamefont {Derzsi}}, \bibinfo {author} {\bibfnamefont {K.}~\bibnamefont {Tokár}}, \bibinfo {author} {\bibfnamefont {P.}~\bibnamefont {Piekarz}},\ and\ \bibinfo {author} {\bibfnamefont {W.}~\bibnamefont {Grochala}},\ }\href@noop {} {\bibfield  {journal} {\bibinfo  {journal} {Phys. Rev. B}\ }\textbf {\bibinfo {volume} {105}},\ \bibinfo {pages} {L081113} (\bibinfo {year} {2022})}\BibitemShut {NoStop}%
\bibitem [{\citenamefont {Dudarev}\ \emph {et~al.}(1998)\citenamefont {Dudarev}, \citenamefont {Botton}, \citenamefont {Savrasov}, \citenamefont {Humphreys},\ and\ \citenamefont {Sutton}}]{413}%
  \BibitemOpen
  \bibfield  {author} {\bibinfo {author} {\bibfnamefont {S.~L.}\ \bibnamefont {Dudarev}}, \bibinfo {author} {\bibfnamefont {G.~A.}\ \bibnamefont {Botton}}, \bibinfo {author} {\bibfnamefont {S.~Y.}\ \bibnamefont {Savrasov}}, \bibinfo {author} {\bibfnamefont {C.~J.}\ \bibnamefont {Humphreys}},\ and\ \bibinfo {author} {\bibfnamefont {A.~P.}\ \bibnamefont {Sutton}},\ }\href@noop {} {\bibfield  {journal} {\bibinfo  {journal} {Phys. Rev. B}\ }\textbf {\bibinfo {volume} {57}},\ \bibinfo {pages} {1505} (\bibinfo {year} {1998})}\BibitemShut {NoStop}%
\bibitem [{\citenamefont {Elcoro}\ \emph {et~al.}(2017)\citenamefont {Elcoro}, \citenamefont {Bradlyn}, \citenamefont {Wang}, \citenamefont {Vergniory}, \citenamefont {Cano}, \citenamefont {Felser}, \citenamefont {Bernevig}, \citenamefont {Orobengoa}, \citenamefont {de~la Flor},\ and\ \citenamefont {Aroyo}}]{139}%
  \BibitemOpen
  \bibfield  {author} {\bibinfo {author} {\bibfnamefont {L.}~\bibnamefont {Elcoro}}, \bibinfo {author} {\bibfnamefont {B.}~\bibnamefont {Bradlyn}}, \bibinfo {author} {\bibfnamefont {Z.}~\bibnamefont {Wang}}, \bibinfo {author} {\bibfnamefont {M.~G.}\ \bibnamefont {Vergniory}}, \bibinfo {author} {\bibfnamefont {J.}~\bibnamefont {Cano}}, \bibinfo {author} {\bibfnamefont {C.}~\bibnamefont {Felser}}, \bibinfo {author} {\bibfnamefont {B.~A.}\ \bibnamefont {Bernevig}}, \bibinfo {author} {\bibfnamefont {D.}~\bibnamefont {Orobengoa}}, \bibinfo {author} {\bibfnamefont {G.}~\bibnamefont {de~la Flor}},\ and\ \bibinfo {author} {\bibfnamefont {M.~I.}\ \bibnamefont {Aroyo}},\ }\href@noop {} {\bibfield  {journal} {\bibinfo  {journal} {J. Appl. Crystallogr.}\ }\textbf {\bibinfo {volume} {50}},\ \bibinfo {pages} {1457} (\bibinfo {year} {2017})}\BibitemShut {NoStop}%
\bibitem [{\citenamefont {Fredericks}\ \emph {et~al.}(2021)\citenamefont {Fredericks}, \citenamefont {Parrish}, \citenamefont {Sayre},\ and\ \citenamefont {Zhu}}]{591}%
  \BibitemOpen
  \bibfield  {author} {\bibinfo {author} {\bibfnamefont {S.}~\bibnamefont {Fredericks}}, \bibinfo {author} {\bibfnamefont {K.}~\bibnamefont {Parrish}}, \bibinfo {author} {\bibfnamefont {D.}~\bibnamefont {Sayre}},\ and\ \bibinfo {author} {\bibfnamefont {Q.}~\bibnamefont {Zhu}},\ }\href@noop {} {\bibfield  {journal} {\bibinfo  {journal} {Comput. Phys. Commun.}\ }\textbf {\bibinfo {volume} {261}},\ \bibinfo {pages} {107810} (\bibinfo {year} {2021})}\BibitemShut {NoStop}%
\bibitem [{\citenamefont {Fu}\ \emph {et~al.}(2024)\citenamefont {Fu}, \citenamefont {Kuisma}, \citenamefont {Larsen}, \citenamefont {Shinohara}, \citenamefont {Togo},\ and\ \citenamefont {Thygesen}}]{618}%
  \BibitemOpen
  \bibfield  {author} {\bibinfo {author} {\bibfnamefont {J.}~\bibnamefont {Fu}}, \bibinfo {author} {\bibfnamefont {M.}~\bibnamefont {Kuisma}}, \bibinfo {author} {\bibfnamefont {A.~H.}\ \bibnamefont {Larsen}}, \bibinfo {author} {\bibfnamefont {K.}~\bibnamefont {Shinohara}}, \bibinfo {author} {\bibfnamefont {A.}~\bibnamefont {Togo}},\ and\ \bibinfo {author} {\bibfnamefont {K.~S.}\ \bibnamefont {Thygesen}},\ }\href@noop {} {\bibfield  {journal} {\bibinfo  {journal} {2D Mater.}\ }\textbf {\bibinfo {volume} {11}},\ \bibinfo {pages} {035009} (\bibinfo {year} {2024})}\BibitemShut {NoStop}%
\bibitem [{\citenamefont {Grimme}\ \emph {et~al.}(2010)\citenamefont {Grimme}, \citenamefont {Antony}, \citenamefont {Ehrlich},\ and\ \citenamefont {Krieg}}]{401}%
  \BibitemOpen
  \bibfield  {author} {\bibinfo {author} {\bibfnamefont {S.}~\bibnamefont {Grimme}}, \bibinfo {author} {\bibfnamefont {J.}~\bibnamefont {Antony}}, \bibinfo {author} {\bibfnamefont {S.}~\bibnamefont {Ehrlich}},\ and\ \bibinfo {author} {\bibfnamefont {H.}~\bibnamefont {Krieg}},\ }\href@noop {} {\bibfield  {journal} {\bibinfo  {journal} {J. Chem. Phys.}\ }\textbf {\bibinfo {volume} {132}},\ \bibinfo {pages} {154104} (\bibinfo {year} {2010})}\BibitemShut {NoStop}%
\bibitem [{\citenamefont {Henkelman}\ \emph {et~al.}(2000)\citenamefont {Henkelman}, \citenamefont {Uberuaga},\ and\ \citenamefont {Jónsson}}]{904}%
  \BibitemOpen
  \bibfield  {author} {\bibinfo {author} {\bibfnamefont {G.}~\bibnamefont {Henkelman}}, \bibinfo {author} {\bibfnamefont {B.~P.}\ \bibnamefont {Uberuaga}},\ and\ \bibinfo {author} {\bibfnamefont {H.}~\bibnamefont {Jónsson}},\ }\href@noop {} {\bibfield  {journal} {\bibinfo  {journal} {The Journal of Chemical Physics}\ }\textbf {\bibinfo {volume} {113}},\ \bibinfo {pages} {9901} (\bibinfo {year} {2000})}\BibitemShut {NoStop}%
\bibitem [{\citenamefont {Hunter}(2007)}]{411}%
  \BibitemOpen
  \bibfield  {author} {\bibinfo {author} {\bibfnamefont {J.~D.}\ \bibnamefont {Hunter}},\ }\href@noop {} {\bibfield  {journal} {\bibinfo  {journal} {Comput. Sci. Eng.}\ }\textbf {\bibinfo {volume} {9}},\ \bibinfo {pages} {90} (\bibinfo {year} {2007})}\BibitemShut {NoStop}%
\bibitem [{\citenamefont {King-Smith}\ and\ \citenamefont {Vanderbilt}(1993)}]{792}%
  \BibitemOpen
  \bibfield  {author} {\bibinfo {author} {\bibfnamefont {R.~D.}\ \bibnamefont {King-Smith}}\ and\ \bibinfo {author} {\bibfnamefont {D.}~\bibnamefont {Vanderbilt}},\ }\href@noop {} {\bibfield  {journal} {\bibinfo  {journal} {Phys Rev B Condens Matter}\ }\textbf {\bibinfo {volume} {47}},\ \bibinfo {pages} {1651} (\bibinfo {year} {1993})}\BibitemShut {NoStop}%
\bibitem [{\citenamefont {Kresse}\ and\ \citenamefont {Furthmüller}(1996)}]{395}%
  \BibitemOpen
  \bibfield  {author} {\bibinfo {author} {\bibfnamefont {G.}~\bibnamefont {Kresse}}\ and\ \bibinfo {author} {\bibfnamefont {J.}~\bibnamefont {Furthmüller}},\ }\href@noop {} {\bibfield  {journal} {\bibinfo  {journal} {Comput. Mater. Sci.}\ }\textbf {\bibinfo {volume} {6}},\ \bibinfo {pages} {15} (\bibinfo {year} {1996})}\BibitemShut {NoStop}%
\bibitem [{\citenamefont {Kresse}\ and\ \citenamefont {Joubert}(1999)}]{394}%
  \BibitemOpen
  \bibfield  {author} {\bibinfo {author} {\bibfnamefont {G.}~\bibnamefont {Kresse}}\ and\ \bibinfo {author} {\bibfnamefont {D.}~\bibnamefont {Joubert}},\ }\href@noop {} {\bibfield  {journal} {\bibinfo  {journal} {Phys. Rev. B}\ }\textbf {\bibinfo {volume} {59}},\ \bibinfo {pages} {1758} (\bibinfo {year} {1999})}\BibitemShut {NoStop}%
\bibitem [{\citenamefont {Momma}\ and\ \citenamefont {Izumi}(2011)}]{410}%
  \BibitemOpen
  \bibfield  {author} {\bibinfo {author} {\bibfnamefont {K.}~\bibnamefont {Momma}}\ and\ \bibinfo {author} {\bibfnamefont {F.}~\bibnamefont {Izumi}},\ }\href@noop {} {\bibfield  {journal} {\bibinfo  {journal} {J. Appl. Crystallogr.}\ }\textbf {\bibinfo {volume} {44}},\ \bibinfo {pages} {1272} (\bibinfo {year} {2011})}\BibitemShut {NoStop}%
\bibitem [{\citenamefont {Ong}\ \emph {et~al.}(2013)\citenamefont {Ong}, \citenamefont {Richards}, \citenamefont {Jain}, \citenamefont {Hautier}, \citenamefont {Kocher}, \citenamefont {Cholia}, \citenamefont {Gunter}, \citenamefont {Chevrier}, \citenamefont {Persson},\ and\ \citenamefont {Ceder}}]{406}%
  \BibitemOpen
  \bibfield  {author} {\bibinfo {author} {\bibfnamefont {S.~P.}\ \bibnamefont {Ong}}, \bibinfo {author} {\bibfnamefont {W.~D.}\ \bibnamefont {Richards}}, \bibinfo {author} {\bibfnamefont {A.}~\bibnamefont {Jain}}, \bibinfo {author} {\bibfnamefont {G.}~\bibnamefont {Hautier}}, \bibinfo {author} {\bibfnamefont {M.}~\bibnamefont {Kocher}}, \bibinfo {author} {\bibfnamefont {S.}~\bibnamefont {Cholia}}, \bibinfo {author} {\bibfnamefont {D.}~\bibnamefont {Gunter}}, \bibinfo {author} {\bibfnamefont {V.~L.}\ \bibnamefont {Chevrier}}, \bibinfo {author} {\bibfnamefont {K.~A.}\ \bibnamefont {Persson}},\ and\ \bibinfo {author} {\bibfnamefont {G.}~\bibnamefont {Ceder}},\ }\href@noop {} {\bibfield  {journal} {\bibinfo  {journal} {Comput. Mater. Sci.}\ }\textbf {\bibinfo {volume} {68}},\ \bibinfo {pages} {314} (\bibinfo {year} {2013})}\BibitemShut {NoStop}%
\bibitem [{\citenamefont {Perdew}\ \emph {et~al.}(1996)\citenamefont {Perdew}, \citenamefont {Burke},\ and\ \citenamefont {Ernzerhof}}]{393}%
  \BibitemOpen
  \bibfield  {author} {\bibinfo {author} {\bibfnamefont {J.~P.}\ \bibnamefont {Perdew}}, \bibinfo {author} {\bibfnamefont {K.}~\bibnamefont {Burke}},\ and\ \bibinfo {author} {\bibfnamefont {M.}~\bibnamefont {Ernzerhof}},\ }\href@noop {} {\bibfield  {journal} {\bibinfo  {journal} {Phys. Rev. Lett.}\ }\textbf {\bibinfo {volume} {77}},\ \bibinfo {pages} {3865} (\bibinfo {year} {1996})}\BibitemShut {NoStop}%
\bibitem [{\citenamefont {Pizzi}\ \emph {et~al.}(2020)\citenamefont {Pizzi}, \citenamefont {Vitale}, \citenamefont {Arita}, \citenamefont {Blügel}, \citenamefont {Freimuth}, \citenamefont {Géranton}, \citenamefont {Gibertini}, \citenamefont {Gresch}, \citenamefont {Johnson}, \citenamefont {Koretsune}, \citenamefont {Ibañez-Azpiroz}, \citenamefont {Lee}, \citenamefont {Lihm}, \citenamefont {Marchand}, \citenamefont {Marrazzo}, \citenamefont {Mokrousov}, \citenamefont {Mustafa}, \citenamefont {Nohara}, \citenamefont {Nomura}, \citenamefont {Paulatto}, \citenamefont {Poncé}, \citenamefont {Ponweiser}, \citenamefont {Qiao}, \citenamefont {Thöle}, \citenamefont {Tsirkin}, \citenamefont {Wierzbowska}, \citenamefont {Marzari}, \citenamefont {Vanderbilt}, \citenamefont {Souza}, \citenamefont {Mostofi},\ and\ \citenamefont {Yates}}]{746}%
  \BibitemOpen
  \bibfield  {author} {\bibinfo {author} {\bibfnamefont {G.}~\bibnamefont {Pizzi}}, \bibinfo {author} {\bibfnamefont {V.}~\bibnamefont {Vitale}}, \bibinfo {author} {\bibfnamefont {R.}~\bibnamefont {Arita}}, \bibinfo {author} {\bibfnamefont {S.}~\bibnamefont {Blügel}}, \bibinfo {author} {\bibfnamefont {F.}~\bibnamefont {Freimuth}}, \bibinfo {author} {\bibfnamefont {G.}~\bibnamefont {Géranton}}, \bibinfo {author} {\bibfnamefont {M.}~\bibnamefont {Gibertini}}, \bibinfo {author} {\bibfnamefont {D.}~\bibnamefont {Gresch}}, \bibinfo {author} {\bibfnamefont {C.}~\bibnamefont {Johnson}}, \bibinfo {author} {\bibfnamefont {T.}~\bibnamefont {Koretsune}}, \bibinfo {author} {\bibfnamefont {J.}~\bibnamefont {Ibañez-Azpiroz}}, \bibinfo {author} {\bibfnamefont {H.}~\bibnamefont {Lee}}, \bibinfo {author} {\bibfnamefont {J.-M.}\ \bibnamefont {Lihm}}, \bibinfo {author} {\bibfnamefont {D.}~\bibnamefont {Marchand}}, \bibinfo {author} {\bibfnamefont {A.}~\bibnamefont {Marrazzo}}, \bibinfo {author} {\bibfnamefont {Y.}~\bibnamefont {Mokrousov}}, \bibinfo {author} {\bibfnamefont {J.~I.}\ \bibnamefont {Mustafa}}, \bibinfo {author} {\bibfnamefont {Y.}~\bibnamefont {Nohara}}, \bibinfo {author} {\bibfnamefont {Y.}~\bibnamefont {Nomura}}, \bibinfo {author} {\bibfnamefont {L.}~\bibnamefont {Paulatto}}, \bibinfo {author} {\bibfnamefont {S.}~\bibnamefont {Poncé}}, \bibinfo {author} {\bibfnamefont {T.}~\bibnamefont {Ponweiser}}, \bibinfo {author} {\bibfnamefont {J.}~\bibnamefont {Qiao}}, \bibinfo {author} {\bibfnamefont {F.}~\bibnamefont {Thöle}}, \bibinfo {author} {\bibfnamefont {S.~S.}\ \bibnamefont {Tsirkin}}, \bibinfo {author} {\bibfnamefont {M.}~\bibnamefont {Wierzbowska}}, \bibinfo {author} {\bibfnamefont {N.}~\bibnamefont {Marzari}}, \bibinfo {author} {\bibfnamefont {D.}~\bibnamefont {Vanderbilt}}, \bibinfo {author} {\bibfnamefont {I.}~\bibnamefont {Souza}}, \bibinfo {author} {\bibfnamefont {A.~A.}\ \bibnamefont {Mostofi}},\ and\ \bibinfo {author} {\bibfnamefont {J.~R.}\ \bibnamefont {Yates}},\ }\href@noop {} {\bibfield  {journal} {\bibinfo  {journal} {J. Phys.: Condens. Matter}\ }\textbf {\bibinfo {volume} {32}},\ \bibinfo {pages} {165902} (\bibinfo {year} {2020})}\BibitemShut {NoStop}%
\bibitem [{\citenamefont {Resta}(1992)}]{905}%
  \BibitemOpen
  \bibfield  {author} {\bibinfo {author} {\bibfnamefont {R.}~\bibnamefont {Resta}},\ }\href@noop {} {\bibfield  {journal} {\bibinfo  {journal} {Ferroelectrics}\ }\textbf {\bibinfo {volume} {136}},\ \bibinfo {pages} {51} (\bibinfo {year} {1992})}\BibitemShut {NoStop}%
\bibitem [{\citenamefont {Resta}(1994)}]{796}%
  \BibitemOpen
  \bibfield  {author} {\bibinfo {author} {\bibfnamefont {R.}~\bibnamefont {Resta}},\ }\href@noop {} {\bibfield  {journal} {\bibinfo  {journal} {Rev. Mod. Phys.}\ }\textbf {\bibinfo {volume} {66}},\ \bibinfo {pages} {899} (\bibinfo {year} {1994})}\BibitemShut {NoStop}%
\bibitem [{\citenamefont {Togo}\ \emph {et~al.}(2024)\citenamefont {Togo}, \citenamefont {Shinohara},\ and\ \citenamefont {Tanaka}}]{1086}%
  \BibitemOpen
  \bibfield  {author} {\bibinfo {author} {\bibfnamefont {A.}~\bibnamefont {Togo}}, \bibinfo {author} {\bibfnamefont {K.}~\bibnamefont {Shinohara}},\ and\ \bibinfo {author} {\bibfnamefont {I.}~\bibnamefont {Tanaka}},\ }\href@noop {} {\bibfield  {journal} {\bibinfo  {journal} {Sci. Technol. Adv. Mater.: Methods}\ }\textbf {\bibinfo {volume} {4}},\ \bibinfo {pages} {2384822} (\bibinfo {year} {2024})}\BibitemShut {NoStop}%
\bibitem [{\citenamefont {Wang}\ \emph {et~al.}(2021)\citenamefont {Wang}, \citenamefont {Xu}, \citenamefont {Liu}, \citenamefont {Tang},\ and\ \citenamefont {Geng}}]{409}%
  \BibitemOpen
  \bibfield  {author} {\bibinfo {author} {\bibfnamefont {V.}~\bibnamefont {Wang}}, \bibinfo {author} {\bibfnamefont {N.}~\bibnamefont {Xu}}, \bibinfo {author} {\bibfnamefont {J.-C.}\ \bibnamefont {Liu}}, \bibinfo {author} {\bibfnamefont {G.}~\bibnamefont {Tang}},\ and\ \bibinfo {author} {\bibfnamefont {W.-T.}\ \bibnamefont {Geng}},\ }\href@noop {} {\bibfield  {journal} {\bibinfo  {journal} {Comput. Phys. Commun.}\ }\textbf {\bibinfo {volume} {267}},\ \bibinfo {pages} {108033} (\bibinfo {year} {2021})}\BibitemShut {NoStop}%
\bibitem [{\citenamefont {Wu}\ \emph {et~al.}(2018)\citenamefont {Wu}, \citenamefont {Zhang}, \citenamefont {Song}, \citenamefont {Troyer},\ and\ \citenamefont {Soluyanov}}]{268}%
  \BibitemOpen
  \bibfield  {author} {\bibinfo {author} {\bibfnamefont {Q.}~\bibnamefont {Wu}}, \bibinfo {author} {\bibfnamefont {S.}~\bibnamefont {Zhang}}, \bibinfo {author} {\bibfnamefont {H.-F.}\ \bibnamefont {Song}}, \bibinfo {author} {\bibfnamefont {M.}~\bibnamefont {Troyer}},\ and\ \bibinfo {author} {\bibfnamefont {A.~A.}\ \bibnamefont {Soluyanov}},\ }\href@noop {} {\bibfield  {journal} {\bibinfo  {journal} {Comput. Phys. Commun.}\ }\textbf {\bibinfo {volume} {224}},\ \bibinfo {pages} {405} (\bibinfo {year} {2018})}\BibitemShut {NoStop}%
\end{thebibliography}

%


\onecolumngrid

\begin{appendix}
\section{End Matter}

\label{appendix}

\emph{Appendix: Generalized stacking rule for compensated ferrimagnets}---In a collinear compensated ferrimagnet (CFiM), the opposite-spin sublattices are not
connected by any symmetry operator. We focus on collinear CFiM bilayers,
which consist of two ferromagnetic monolayers with opposite spins. Since
the two monolayers cannot be connected by a symmetry, CFiM bilayers
require breaking \(R^{-}\)symmetry, and thus they are permitted for all
stacking orders that allow sliding ferroelectricity, as listed in Tables
S1-S3 \cite{1013}. For two CFiM bilayers exhibiting
sliding ferroelectricity, they are connected by \(N^{-}\). Considering
spin rotation, they are connected by
\(N_{s}^{-} = \{ 2_{\bot}||N^{-}\}\) because the top and bottom
monolayers have opposite spins. \(N_{s}^{-} = \{ 2_{\bot}||N^{-}\}\)
enables switching of nonrelativistic spin polarization, indicating that
all considered CFiM bilayers can exhibit sliding ferroelectric control
of spin polarization.

We next consider the SOC effect and discuss the sliding ferroelectric
control of AHE in CFiM bilayers. Specifically, we consider the
ferroelectric CFiM bilayer with an out-of-plane magnetic configuration,
which can host AHE since its symmetry permits the out-of-plane
magnetization and thus permits a nonzero Berry curvature. The AHE can be
switched under ferroelectric switching since the connection operator
\(N_{m}^{-}\) can reverse the out-of-plane magnetization and thus
reverse the Berry curvature. Therefore, for all ferroelectric CFiM
bilayers with out-of-plane magnetic configuration and stacking orders
listed in Tables S1-S3 \cite{1013},  their spin
polarization and AHE can be simultaneously controlled via sliding
ferroelectricity.
 \end{appendix}

\end{document}